\documentclass[preprint2]{aastex631}
\usepackage[caption=false]{subfig}
\usepackage{orcidlink}

\defcitealias{Wetzell2022}{Wet22}

\begin{document}
\title{Probing Projection Effects in Optically-Selected Clusters with Velocity Dispersions Outliers}

\author{K. ~Graham \orcidlink{0009-0002-2369-4722}}
\affiliation{University of California, Santa Cruz, Santa Cruz, CA 95064, USA}
\affiliation{Santa Cruz Institute for Particle Physics, Santa Cruz, CA 95064, USA}
\author{T.~E.~Jeltema \orcidlink{0000-0001-6089-0365}}
\affiliation{University of California, Santa Cruz, Santa Cruz, CA 95064, USA}
\affiliation{Santa Cruz Institute for Particle Physics, Santa Cruz, CA 95064, USA}
\author{C.-H. To \orcidlink{0000-0001-7836-2261}}
\affiliation{Center for Cosmology and Astro-Particle Physics, The Ohio State University, Columbus, OH 43210, USA}
\author{V. Wetzell \orcidlink{0000-0003-0814-2858}}
\affiliation{Department of Physics \& Astronomy
University of Pennsylvania
209 South 33rd Street
Philadelphia, PA 19104-6396}
\author{Kyle W. Davis \orcidlink{0000-0002-5680-4660}}
\affiliation{University of California, Santa Cruz, Santa Cruz, CA 95064, USA}
\author{Conghao Zhou \orcidlink{0000-0002-2897-6326}}
\affiliation{University of California, Santa Cruz, Santa Cruz, CA 95064, USA}
\affiliation{Santa Cruz Institute for Particle Physics, Santa Cruz, CA 95064, USA}
\author{H.-Y. Wu \orcidlink{0000-0002-7904-1707}}
\affiliation{Department of Physics, Southern Methodist University, Dallas, TX 75205, USA}
\author{E.S. Rykoff \orcidlink{0000-0001-9376-3135}}
\affiliation{Kavli Institute for Particle Astrophysics and Cosmology, P. O. Box 2450, Stanford University, Stanford, CA 94305, USA}
\affiliation{SLAC National Accelerator Laboratory, Menlo Park, CA 94025, USA}
\author{J. DeRose \orcidlink{0000-0002-0728-0960}}
\affiliation{Lawrence Berkeley National Laboratory, 1 Cyclotron Road, Berkeley, CA 94720, USA}
\author{Michel Aguena \orcidlink{0000-0001-5679-6747}}
\affiliation{Université Paris Cité, CNRS(/IN2P3), Astroparticule et Cosmologie, F-75013 Paris, France}
\author{R.H. Wechsler \orcidlink{0000-0003-2229-011X}}
\affiliation{Kavli Institute for Particle Astrophysics and Cosmology, P.O.~Box 2450, Stanford University, Stanford, CA 94305, USA}

\begin{abstract}
The evolution of galaxy cluster abundance is a powerful probe of cosmology, but its efficacy depends on our ability to understand and control for systematics in the selection and mass estimation of clusters.  
We study redMaPPer-selected clusters in the Cardinal simulation, which simulates the DES Y3 observational conditions. 
Specifically, we investigate the line-of-sight velocity distributions of cluster member galaxies in 12,272 simulated clusters. We find that a significant fraction of clusters, 35\%, have velocity dispersions that are high compared to their richness with an offset of $\sim 400$ km s$^{-1}$ between the outliers and the main population. 
A similar population of velocity dispersion outliers was found in the DES Y3 data. 
Nearly all of the outliers (98\%) in a low-richness subsample have non-Gaussian velocity distributions, and they are more likely than the main population to have significant velocity substructure. 
These clusters also tend to have higher redshifts than clusters with typical velocity dispersions, and they show a larger scatter between halo mass and richness.
Overall, our results point to a significant contamination of redMaPPer-selected samples from line of sight structure, particularly at high redshifts and low richness.  Comparison to the Buzzard simulations shows similar results.
\end{abstract}

\section{Introduction}
Galaxy clusters offer an excellent probe into the nature of dark energy and dark matter as cluster structure and abundance are highly sensitive to the universe’s expansion rate and mass density, thus, the underlying cosmology \citep[e.g.][]{Allen2011, Kravtsov2012}. 
In fact, clusters are one of the corner stone probes for large-area imaging surveys like the Dark Energy Survey (DES) and the Legacy Survey of Space and Time (LSST).  Theoretically, the very large numbers of clusters over a broad mass range in these surveys should allow for among the tightest constraints on cosmology \citep{Weinberg2013, DES2020}. 

Unfortunately, optical, photometric cluster selection is prone to systematic biases, in particular due to projection effects.  The inability in imaging surveys to distinguish structure along the line of sight can bias both the selection of clusters and their observed richnesses, used to estimate cluster mass \citep[e.g.][]{Lucey1983, Rozo2015b, Sohn2018, Costanzi2019, Sunayama20, Myles2021, Wu22, Zhou23}.  In particular, the DES Y1 results imply unmodeled systematics for low richness clusters selected by the redMaPPer algorithm \citep{Rykoff2014, Rykoff2016}, whose large numbers provide much of the potential cosmological constraining power \citep{DES2020}.

In simulations, it has been found that the red-sequence-based selection employed by the redMaPPer is biased toward the selection of lines of sight with boosted lensing signals compared to a purely mass-based selection \citep{Sunayama20, Wu22, Nde26}. This bias is likely due to the selection of clusters in regions with filaments or groups along the line of sight leading to a correlated scatter between richness and lensing. Simulations can be used to calibrate for these selection effects if the simulations appropriately reflect the real data.  For example, \cite{Costanzi26} presents a forward modeling approach to account for projection effects given constraints from simulations or multiwavelength data. \cite{Zhou23} found using simulations that the selection bias in the stacked weak lensing signal could be modeled and corrected for based on the comparison of the stacked lensing profiles for Sunyaev-Zel'dovich effect (SZ) detected and undetected redMaPPer clusters, giving a potential path forward in constraining the average bias using observations once other effects like miscetnering are accounted for. 

In observations, \cite{Grandis21} found based on the detection fraction of redMaPPer clusters in South Pole Telescope (SPT) SZ selection that fewer redMaPPer clusters were detected in SZ than expected at low richness implying a scatter or contamination that grows at lower richnesses.  However, the SZ-detected samples are generally limited to higher mass clusters.  \cite{Kelly24} did not find a significant increase in the scatter between X-ray temperature and redMaPPer richness, though the X-ray sample sizes limited this analysis to two richness bins. Using stacked SDSS and DESI spectroscopy \cite{Myles2021} and \cite{myles25} were able to constrain the average projection contribution to observed richness.  These works shows that the bias in observed richness due to projection grows for decreasing richness and increasing redshift, but this work only constrains the average effect and at low redshifts. 

While a lot of progress has been made in understanding projection and selection biases, questions remain in terms of the size of these biases, their correlation with richness and redshift, and to what extent simulations accurately predict these effects.  Is it projection effects along a fraction of lines of sight contributing to selection bias or simply a larger scatter in observed richness compared to halo mass at lower richnesses?

In this paper, we follow up on \cite{Wetzell2022} (hereafter: \citetalias{Wetzell2022}) which calculated the velocity dispersion for individual clusters with sufficient spectroscopy for redMaPPer-selected clusters from the the Year 3 Gold catalog of the Dark Energy Survey \citep{Sevilla-Noarbe2021}. This paper found a population of high velocity dispersion, low richness outlier clusters. These outliers were found to be more frequent at high redshifts and lower richnesses and could indicate contamination of the cluster catalog by systems with significant line-of-sight structure \citepalias{Wetzell2022}. In order to do better understand the origin of these systems, we search for the same outlier clusters in simulations, specifically the Cardinal mock galaxy catalogs \citep{To2023} and investigate their properties. We look at 13234 mock clusters selected by redMaPPer run on the Cardinal simulations and investigate their velocity dispersion-richness relation and velocity structure. 
We also investigate the halo mass of these clusters and their relation to the outliers, extending the work of \citetalias{Wetzell2022}. 

This paper is structured as follows. In Section 2, we present our cluster selection and an overview of the methods used to obtain the velocity dispersions. In Section 3, we discuss the velocity dispersion-richness relation and further investigate the properties of the outlier clusters by looking at other observables such as redshift and halo mass. In Section 4, we compare our results to that of another simulation, Buzzard \citep{derose2019}. In Section 5, we summarize the results and discuss future work.

\section{Data \& Methodology}
\subsection{Catalogs}
We study the properties of clusters selected from the Cardinal mock galaxy catalogs, meant to support large scale surveys such as DES, LSST, and DESI \citep{To2023}. The catalogs are taken from a one quarter sky simulation with galaxies populated up to redshift of z = 2.35 and a depth of ${m}_{\tau}$= 27. Compared to older mock galaxy simulations, Cardinal improves upon the subhalo abundance model and color assignment model to better model galaxy clustering. 

Similar to the Dark Energy Survey, the redMaPPer algorithm, a photometric, red-sequence cluster finder \citep{Rykoff2014, Rykoff2016, To2023} was run on the Cardinal simulations to identify clusters and their member galaxies. RedMaPPer uses spectroscopic data to generate a red-sequence template with which it can determine cluster redshifts. After selecting bright and red galaxies as potential cluster centers, redMaPPer iteratively determines the probability of membership to a cluster of the surrounding galaxies in the field by comparing their spatial separation, magnitude, and color to the most likely central cluster galaxy. Finally, redMaPPer determines a richness ($\lambda$) based on the sum of member galaxies’ probabilities. Richness serves as a good observable cluster mass proxy due to its strong correlation to halo mass \citep{Rozo&Rykoff2014, Rozo2015a, McClintock2018, Abbott2019, To2021}.

The data set employed in this paper uses galaxy clusters and their member galaxies selected by redMaPPer version 0.8.1 (python version) in Cardinal 3Y6a version 2.0. In particular, we use the richness greater than 20, full cluster catalog with its associated member catalog. We focus on a subset of clusters with sufficient member data after a velocity offset cut as described in the following section. This leaves us with a final sample of 12,272 clusters for our analysis. 

It should be noted that in the Cardinal simulations the galaxies inherit the velocities of the underlying dark matter particles to which they are assigned.  As the dynamics of cluster galaxies can be biased relative to the underlying dark matter \citep[e.g.][]{Wu2013}, our velocity dispersion measurements based on dark matter particles may not explicitly match what would be found observationally.  However, the bias is small ($\sim 10$\%) and depends on the brightness of galaxies used \citep[e.g.][]{Wu2013}.  In addition, when defining cluster populations based on velocity dispersion, we do so self-consistently from the simulations alone.

\subsubsection{Richness and Redshift Distributions of Catalogs}
\begin{figure*}
    \includegraphics[width = 0.49\linewidth]{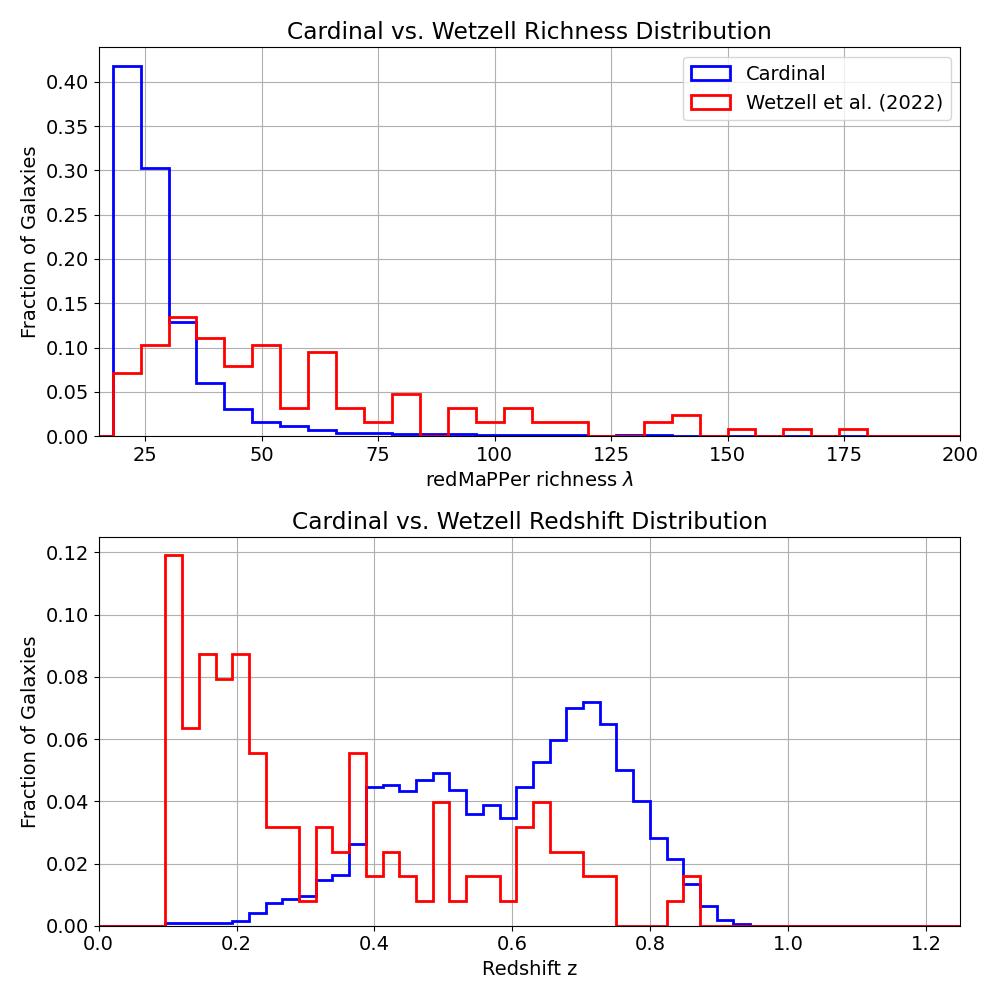}
    \includegraphics[width = 0.49\linewidth]{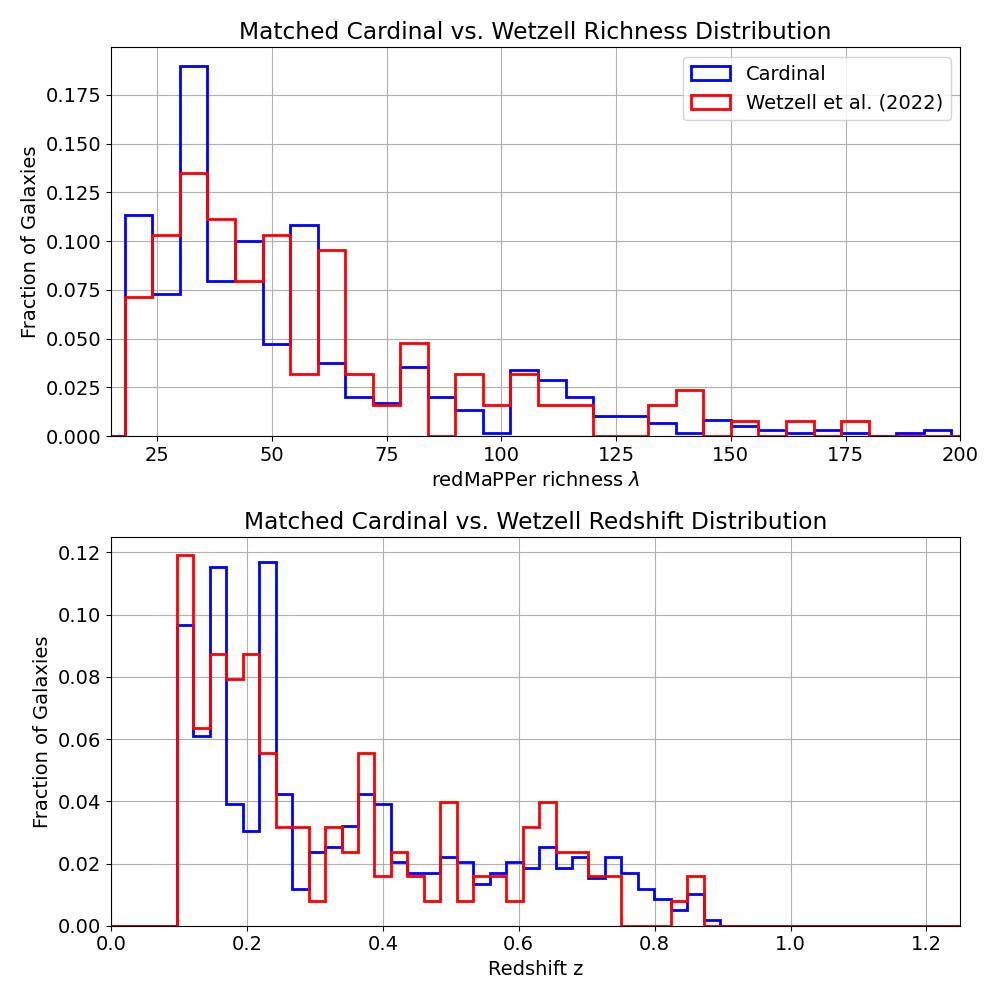}
    \caption{Histograms of the distribution of richness and redshift for the Cardinal clusters (blue) and the clusters from \citetalias{Wetzell2022} (red). The graphs on the left side have the full Cardinal catalog while the graphs on the right side have the matched Cardinal sample.}
    \label{fig:match_verndist}
\end{figure*}

In addition to exploring the properties of redMaPPer clusters using simulations, in this paper we compare our results to the observational results of \citetalias{Wetzell2022}.  We would also like to understand from the simulations the expectations for the full DES cluster catalogs.  In making these comparisons, it is important to understand the 
relative distributions in observable properties, such as richness and redshift, between catalogs. In particular, the sample of \citetalias{Wetzell2022} is limited to clusters with sufficient archival spectroscopy, giving a relatively small sample and one whose richnesses and redshifts do not necessarily reflect the overall cluster population.  Given that the outlier clusters tend to be low richness and tend to have higher redshifts, this discrepancy may affect the resultant outlier fraction. The plots on the left side of Figure~\ref{fig:match_verndist} show the distribution of richness and redshift of our Cardinal clusters compared to that of the clusters used in \citetalias{Wetzell2022}. 

Qualitatively, it is clear that neither the redshifts nor the richnesses are distributed similarly between these catalogs. In particular, our sample contains a higher fraction of clusters at lower richnesses indicating that we might find a larger outlier fraction than \citetalias{Wetzell2022}. In order to better compare outlier fractions, we randomly select clusters from our sample to match the distribution of those in \citetalias{Wetzell2022}. To do this, we split both catalogs into evenly spaced bins of redshift in steps of 0.1 and bins of richness spaced to give smaller bins at low richnesses where the sample size is larger (i.e. the edges of the richness bins are 20, 30, 40, 50, 70, 100, 500). We then select clusters that fall within both a given richness and redshift bin for both catalogs. We use the number of clusters selected from \citetalias{Wetzell2022} times a constant factor, which we chose to be 5, to determine how many clusters we randomly draw from the selected Cardinal sample. Doing this process over all bins yields a ``matched" Cardinal sample that should have a similar distribution to the \citetalias{Wetzell2022} catalog. The graphs on the right side of Figure~\ref{fig:match_verndist} shows the distributions of richness and redshift of the matched Cardinal sample with those from \citetalias{Wetzell2022}.

To quantify how well the matching has done, we performed a two-sample Kolmogorov-Smirnov Test (KS Test). The KS test between the richness distributions returned a p-value of $0.52$ and a p-value of $0.28$ for the redshift distributions. These p-values indicate that there is no significant difference between the richness and redshift distributions of the matched Cardinal sample and the clusters from \citetalias{Wetzell2022}. Thus, when comparing outlier fractions, we will utilize the matched Cardinal sample. 

\begin{figure*}
    \includegraphics[width = .49\linewidth]{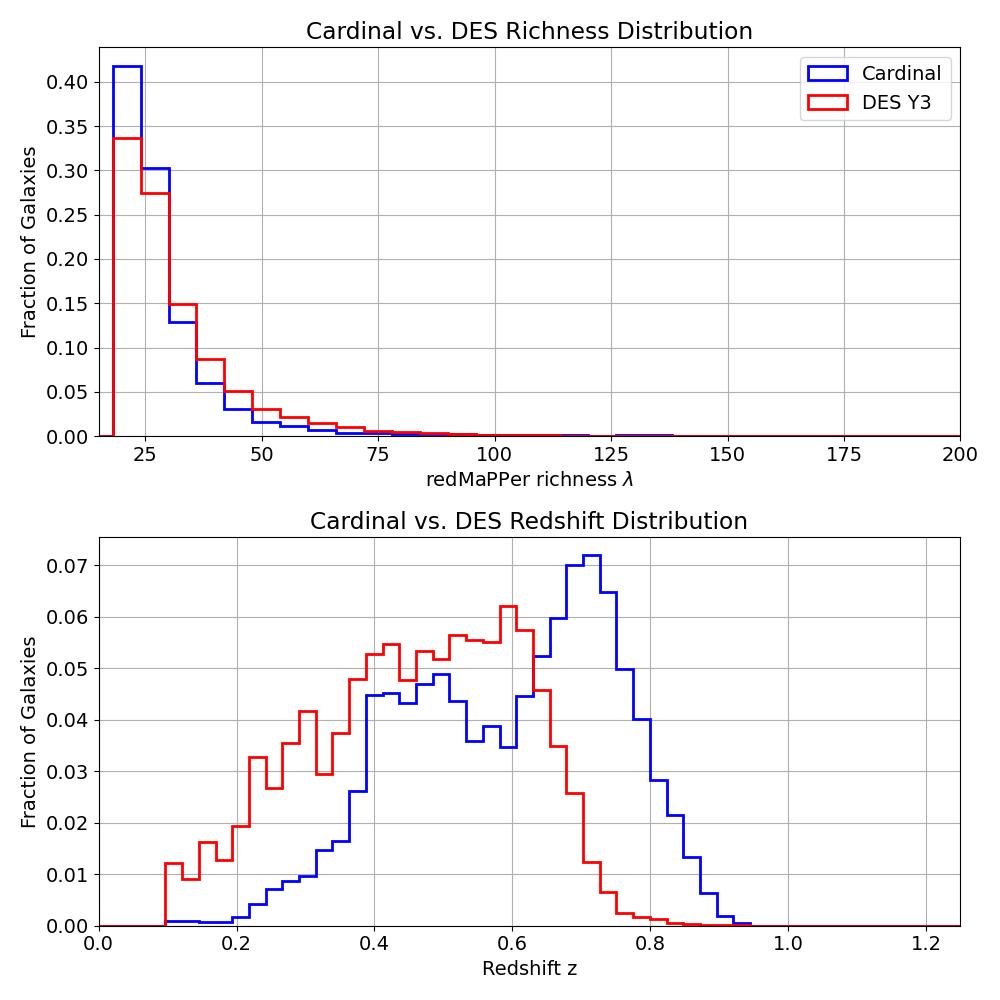}
    \includegraphics[width = .49\linewidth]{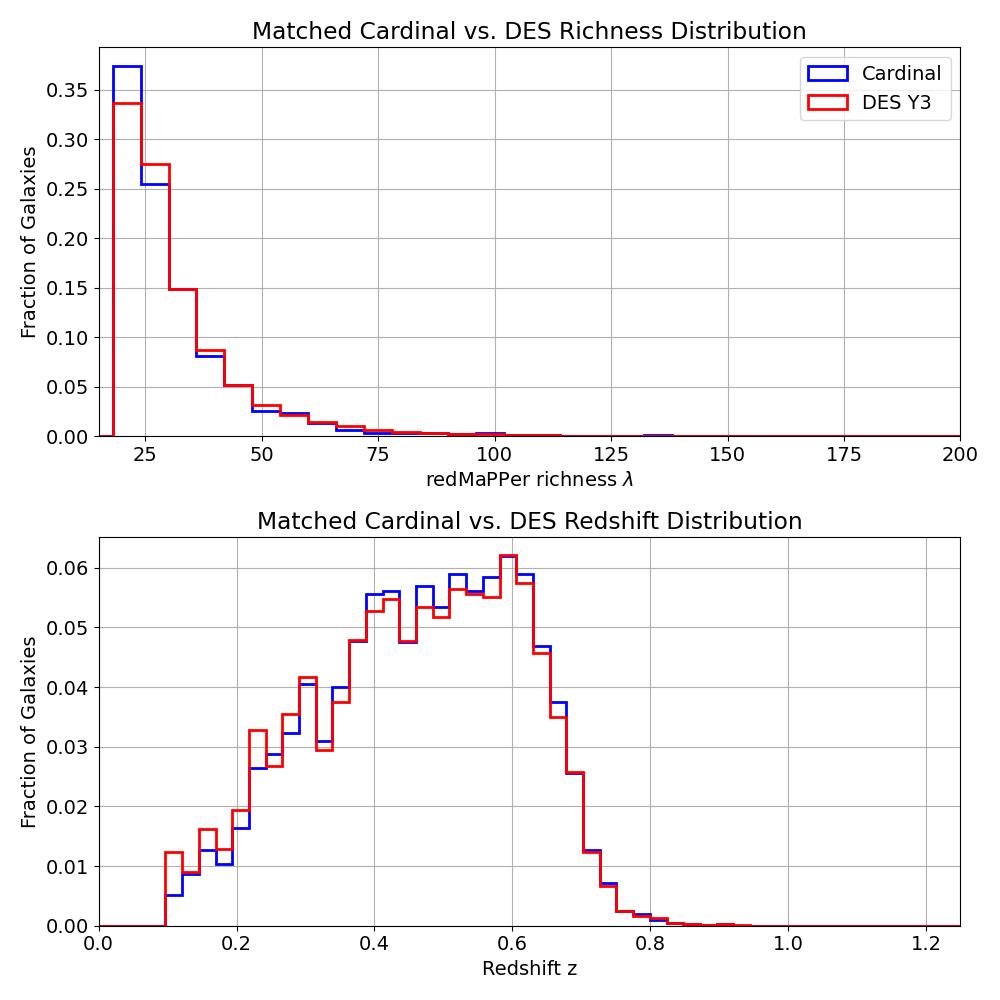}
    \caption{Histograms of the distribution of richness and redshift for the Cardinal clusters (blue) and the clusters from the DES Y3 volume-limited catalog (red). The graphs on the left side have the full Cardinal catalog while the graphs on the right side have the matched Cardinal sample. }
    \label{fig:match_desdist}
\end{figure*}

However, the catalog from \citetalias{Wetzell2022} is not representative of the selection present in the DES catalog. Therefore, in order to understand the expected outlier fraction for a population representative of the DES catalog, we compared the Cardinal sample to the Y3 DES volume-limited catalog as shown in Figure~\ref{fig:match_desdist}. While the Cardinal clusters have a reasonably similar richness distribution to the DES Y3 catalog, the redshift distributions are not well matched. We performed the same matching process except that we chose not to include the constant factor of 5 when determining how many clusters to draw for the matched sample given the much larger size of the DES cluster catalog. The plots on the right side of Figure~\ref{fig:match_desdist} show that the re-sampled Cardinal distribution and the DES Y3 catalog are well-matched. We will discuss the outlier fractions of these matched samples in Section 3.

\subsection{Methodology for Determining Velocity Dispersion}
For determination of velocity dispersions and cluster central redshifts, we use the methods outlined in \citetalias{Wetzell2022}, themselves detailed in \cite{Beers1990}. Specifically, we use the biweight location estimator and gapper scale estimator which perform well for few $N_{Members}$ \citep{Beers1990}. We forgo using the biweight scale estimator used in \citetalias{Wetzell2022} due to it being in good agreement with the gapper method (see Figure 2 of \citetalias{Wetzell2022}). Additionally, the gapper method returns a nearly constant estimate of the velocity dispersion regardless of the number of sampled galaxies in comparison to the biweight scale estimator which performs worse at lower $N_{Members}$ \citep{Ferragamo2020}.

\subsubsection{Biweight Location Estimator}
We use the biweight location estimator to determine the cluster central redshifts based on the mock redshifts of member galaxies. We choose this estimator for the same reasons as \citetalias{Wetzell2022}, that being its robustness for non-Gaussian distributions and its resistance to contaminated normal distributions.

For a set of redshift measurements $\{z_i\}$, the biweight location estimator is defined as
\begin{equation}
    C_{BI}(Z) = M + \frac{\sum_{|u_{i}| < 1}(z_{i} - M)(1 - u_{i}^{2})^{2}}{\sum_{|u_{i}| < 1}(1 - u_{i}^{2})^{2}}
\end{equation}
where M is the median of the sample and $u_{i}$ is defined as
\begin{equation}
    u_{i} = \frac{(z_{i} - M)}{C\,{\rm MAD}(z_{i})}
\end{equation}
$C$ is a “tuning constant” which is chosen as $C = 6.0$ to give $C_{BI}$ a high efficiency for a broad range of initial distributions, and MAD is the median absolute deviation of the galaxy spectroscopic redshifts defined by
\begin{equation}
    {\rm MAD} = \mathrm{median}(|z_{i} - M|)
\end{equation}

As discussed in both \cite{Beers1990} and \citetalias{Wetzell2022}, we iterate through this process 10 times by setting $M$ equal to $C_{BI}$ in order to obtain a more accurate result for the cluster central redshift.

\subsubsection{Gapper Method}
The gapper method is a velocity dispersion estimator based on the gaps between ordered statistics. For ordered measurements of the form $v_{i},v_{i+1}, ..., v_{n}$, with gaps defined by
\begin{equation}
    g_{i} = v_{i+1} - v_{i},\qquad i = 1,2, ..., n-1
\end{equation}
and a set of approximately Gaussian weights defined by
\begin{equation}
    w_{i} = i(n-i)
\end{equation}
the gapper scale estimator can be found using
\begin{equation}
    \sigma_{G} = \frac{\sqrt{\pi}}{n(n-1)} \sum_{i = 1}^{n - 1}w_{i}g_{i}.
\end{equation}

As mentioned in \citetalias{Wetzell2022}, the gapper method is well suited for our data as it gives accurate scale estimates for as few as $N_{Members} = 10$ without being severely affected by interlopers. 

\subsection{Methodology for Cluster and Member Selection}
In this section, we summarize the selection methods used on the Cardinal cluster and member catalogs.  These cuts match those used for DES Y3 clusters in \citetalias{Wetzell2022}, and more details can be found in that work.

In order to robustly probe cluster membership and to measure the peculiar velocity distributions, we limit our sample of clusters to those with at least 15 galaxies after the cuts discussed here. 

Using the cluster central redshift determined by the biweight location estimator, the peculiar velocities for potential member galaxies can be determined using the following equation,

\begin{equation}
v = c \frac{z_{i} - C_{BI}}{1+C_{BI}}
\end{equation}

where c is the speed of light in km s$^{-1}$, $z_{i}$ is the spectroscopic redshift of the galaxy, and $C_{BI}$ is the cluster central redshift determined using the biweight location estimator.

\begin{figure}
    \includegraphics[width=0.98\linewidth]{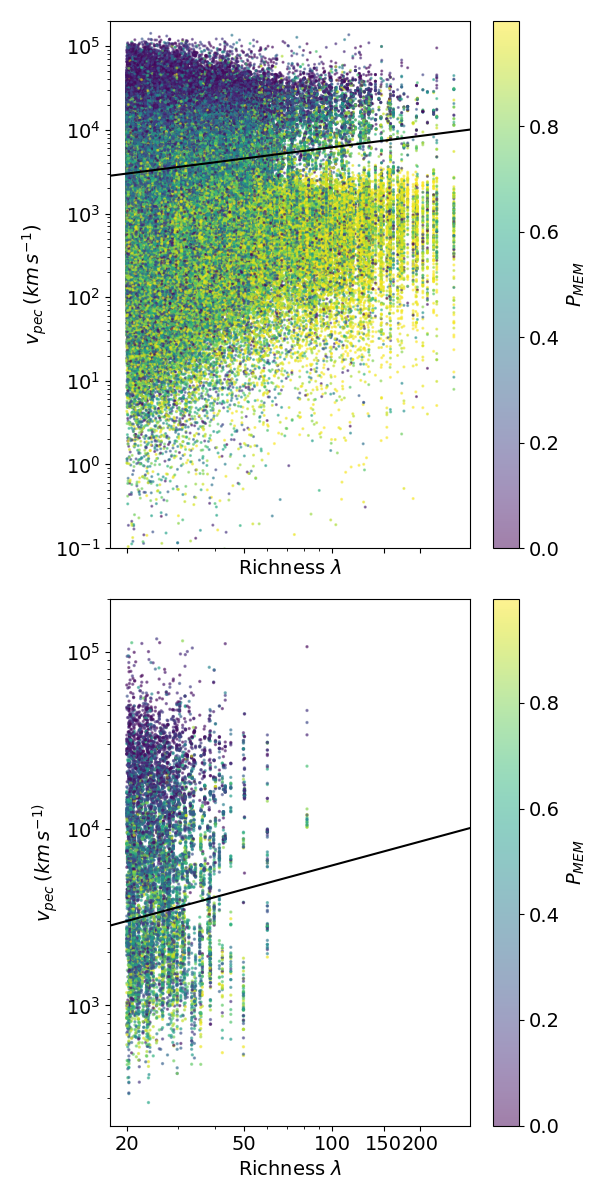}
    \caption{\textit{Top:} The peculiar velocity-richness relation of putative redMaPPer member galaxies color coded by $P_{MEM}$. The black line denotes the relation $v_{pec} = (3000 \mathrm{km s}^{-1})(\frac{\lambda}{20})^{0.45}$. This relation was used as the initial cut to remove interlopers. \textit{Bottom:} The peculiar velocity-richness relation of member galaxies for clusters with too few galaxies to compute velocity dispersion. Colors and black line are the same as for the top panel. }
    \label{fig:velcut}
\end{figure}

After finding the peculiar velocities of redMaPPer potential members, we cut galaxies with large velocity offsets in order to remove interlopers that are not cluster members. We utilize the same richness-dependent cut as \cite{Rozo2015b} and \citetalias{Wetzell2022}:

\begin{equation}
|v| \leq (3000 \mathrm{km/s})\left(\frac{\lambda}{20}\right)^{0.45}
\end{equation}

where $\lambda$ is the richness of the cluster to which redMaPPer assigns a given galaxy as a member. The top panel of Figure~\ref{fig:velcut} presents galaxy peculiar velocity versus cluster richness for our uncut sample with galaxies above the line being cut from the sample; the color coding indicates the galaxy membership probability ($P_{MEM}$). 

After the velocity offset cut, some clusters are left with fewer than 15 members, and a velocity dispersion is not computed.  The bottom panel of Figure~\ref{fig:velcut} shows peculiar velocity versus cluster richness for these clusters. As expected, they tend to be lower richness clusters with fewer potential member galaxies to start with. The clusters for which we do not compute a velocity dispersion make up approximately 3\% of all clusters in our sample indicating that our choice of $N_{Members} = 15$ does not significantly reduce the sample size.  However, these clusters do indicate a population of clusters where projection significantly affects the observed cluster richness.

\subsection{Confidence Intervals}
Uncertainties in the velocity dispersion estimates were determined using a bootstrap resampling following \citetalias{Wetzell2022}; starting from the initial set of cluster member galaxies we draw a new sample of peculiar velocities by randomly drawing from the initial sample with replacement to generate a new sample of the same size. For each cluster, we created 1000 resampled galaxy catalogs. We then applied the gapper method to the resampled cluster velocity distributions to estimate the range of plausible velocity dispersions. We chose the median of these resampled dispersions to be our nominal dispersion for the cluster, and we define a confidence interval based on the range of values spanned by the central 68\% of the bootstrap measurements about the median. We chose this method of determining the confidence intervals due to its lack of assumption of the form of the sampling distribution \citep{Beers1990}.

\subsection{Gaussianity and Substructure Tests}
When evaluating the peculiar velocity distributions for each cluster, we utilize two statistical tests to quantify the Gaussianity of the distribution and the presence of substructure. The tests we used are the Anderson-Darling Test and the Dressler-Shectman Test, respectively.

\subsubsection{Anderson-Darling Test}

To test the Gaussianity of the velocity distributions quantitatively, we implemented the Anderson-Darling (AD) goodness-of-fit test. We use the AD statistic to compare the cumulative distribution functions (CDF) of the galaxy velocities to a Gaussian distribution as calculated below:

\begin{equation}
    A^{2} = -n - \frac{1}{n}\sum^{n}_{i=1} (2i-1)(ln \Phi (x_{i}) + ln(1-\Phi(x_{n+1-i}))) 
\end{equation}

\begin{equation}
    A^{*2} = A^{2}(1+\frac{0.75}{n}+\frac{2.25}{n^{2}})
\end{equation}

where $\Phi(x_{i}), x_{i}\leq x \leq x_{i+1}$ is the CDF of a Gaussian distribution \citep{DAgostino1986}. The p-values, the probability of obtaining $A^{*2}$ given a Gaussian distribution, are then calculated using the equation:

\begin{equation}
    p = a \ \mathrm{exp}(-A^{*2}/b)    
\end{equation}
where $a = 3.6789468$ and $b = 0.1749916$ \citep{Nelson1998}. The velocity distributions are determined to be non-Gaussian if the significance is greater than the 95 percent confidence level ($p-\mathrm{value} < 0.05$) using this test. 

\subsubsection{Dressler-Shectman Test}

We also test for the presence of substructure by applying the Dressler-Shectman (DS) test to our velocity distributions \citep{DS1988}. This statistic compares the local mean velocity and velocity dispersion for subgroups of galaxies to the overall cluster velocity and dispersion, and it is high for clusters with kinematically distinct substructures. For each member galaxy $i$ of a given cluster, we find the local mean velocity ($\overline{v^{i}_{local}}$) and the local velocity dispersion ($\sigma^{i}_{G, local}$) using the $i$th galaxy plus a number of its nearest neighbors ($N_{nn}$). We can then compute:

\begin{equation}
    \delta_{i} = (\frac{N_{nn}+1}{\sigma^{2}_{G}})[(\overline{v^{i}_{local}}-\overline{v})^{2} + (\sigma^{i}_{G, local} - \sigma_{G})^{2}]
\end{equation}

where $1 \leq i \leq n_\mathrm{members}$, $\overline{v}$ is the mean cluster velocity, $\sigma_{G}$ is the cluster velocity dispersion, and $N_{nn} = \sqrt{n_\mathrm{members}}$ rounded down to the nearest integer. The DS statistic can then be calculated as:

\begin{equation}
    \Delta = \sum^{n}_{i=1} \delta_{i}
\end{equation}

We follow \cite{Hou2013} in determining the p-value of the DS test. This is done by comparing the observed $\Delta$ values to shuffled $\Delta$ values which are found by randomly shuffling the galaxy velocities and re-assigning them to galaxy positions. The p-value is then defined to be the number of shuffled $\Delta$'s which are higher than the observed $\Delta$ divided by the number of shuffled realizations of the data
for which we use $n_\mathrm{shuffle} = 100$. We define a cluster to have significant substructure if it has a $p-\mathrm{value} < 0.05$, that is less than 5\% of the randomly shuffled samples have $\Delta$ as high as the observed $\Delta$.  Note that for less than 20 members, the DS test should be taken as a lower limit on the percentage of clusters with substructure \citep{Hou2013}.

\subsection{Halo Matching and Membership}
\label{subsec:halo_match}

In this work, we use clusters selected by redMaPPer run on the Cardinal galaxy catalogs without knowledge of the underlying dark matter halos. This means that there is not a one-to-one correspondence between redMaPPer clusters and halos, and their positions may differ.  In order to compare our cluster properties to the underlying halos, we match clusters and halos using the ClEvaR code\footnote{https://github.com/LSSTDESC/ClEvaR}.  The matching is done based on proximity in projected distance and redshift with a match radius of $R\leq2$ Mpc and a redshift offset of $\Delta z\leq0.05$, similar to the criteria used in \citet{Zhang23}. Here we match clusters to main halos only excluding subhalos. In the case of multiple matches preference was given to the more massive halo matched to a given cluster; matching in the other direction, clusters to halos, yielded the same match in 99.9\% of cases. The median offset between the cluster-halo pairs in redshift and the median physical separation between the pairs are -0.006 and 50 kpc, respectively. These values are small and indicate that the matching parameters are reasonable.

For a given central halo, we can also identify which Cardinal galaxies belong to that halo.  In order to directly compare to redMaPPer's member selection, which includes luminosity and red-sequence color cuts, we start with galaxies in the redMaPPer Cardinal member catalog.  The member catalog includes all galaxies with a redMaPPer membership probability of at least 1\%.  The Cardinal Gold catalog lists the nearest halo to each galaxy and the distance to that halo; galaxies within $R_{200m}$ of a given halo are taken to be members of that halo.  Starting from the redMaPPer member catalog, we identify those galaxies within $R_{200m}$ of their nearest halo and assign them as members of that halo; if the nearest halo is a subhalo then the galaxy is assigned as a member of the host halo.  The ``true" richness of a cluster is then the sum of the number of member galaxies in the halo that was matched to a given cluster.  Note that this is not the true number of galaxies in a halo as we have limited ourselves to galaxies chosen by redMaPPer which are red-sequence and of sufficient luminosity.

\section{Results}
\subsection{Velocity Dispersion-Richness Relation}
\begin{figure*}
    \includegraphics[width = 1\linewidth]{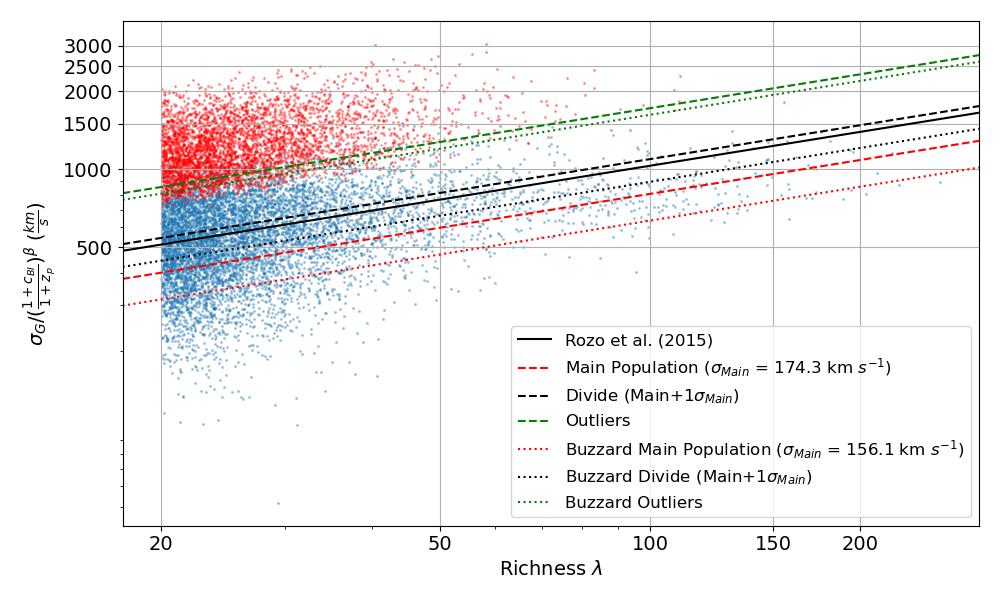}
    \caption{Velocity dispersion-richness relation for redMaPPer-selected clusters in the Cardinal simulations with velocity dispersion estimated via the gapper method. The redshift dependence is accounted for by adjusting the cluster velocity dispersions based on cluster redshift ($z_{p} = 0.171, \beta = 0.54$) following \cite{Rozo2015b}. The solid black line shows the $\lambda-\sigma_{G}$ trend line found by \cite{Rozo2015b}. The population lines were found by fitting a double Gaussian to the residuals relative to the trend line from \cite{Rozo2015b}. The red dashed line is located at the center of the main population. The green dashed line is located at the center of the outlier population. The black dashed line is 1$\sigma$ away from the main population. The dotted lines of the same color correspond to the same locations, but for the Buzzard simulation catalog. Outliers were defined as clusters with lower velocity dispersion limits above the black dashed line. Outlier clusters are marked in red, while the main population is shown in blue.}
    \label{fig:veldisp_richness}
\end{figure*}

In total, we estimate the velocity dispersion using the gapper method for 12,272  simulated clusters.  The relationship between redMaPPer cluster richness and velocity dispersion is shown in Figure~\ref{fig:veldisp_richness}. 

Figure~\ref{fig:veldisp_richness} reveals a bimodal distribution in the velocity dispersion-richness relation similar to what was seen in \citetalias{Wetzell2022}, though in the simulations there is more of a continuum between the two populations.
In general, the $\sigma_G-\lambda$ relation follows a power law with a slope similar to that seen in observed clusters of $\sim0.44$ ($\sigma_{G}$ scales as $\sim0.44$) \citep[e.g.][]{Rozo2015b}. However, there is a large scatter, especially at low richness.  In particular, as was seen in DES Y3 clusters, we find a significant population of clusters with velocity dispersions which are high for their richnesses analogous to the outlier population 
seen in \citetalias{Wetzell2022}.

\begin{figure}
\includegraphics[width = 0.98\linewidth]{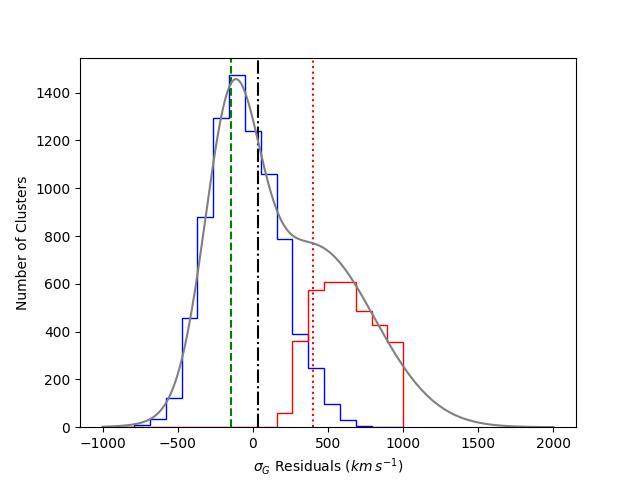}
\caption{Histogram of the residuals between the $\sigma-\lambda$ relation found in \cite{Rozo2015b} and the simulated cluster velocity dispersions, which are the medians of the bootstrap samples. The primary peak is centered at $-132$ km s$^{-1}$ (green dashed line) and has a one sigma-width of $189$ km s$^{-1}$. The secondary peak due to the outlier population is centered at $413$ km s$^{-1}$ (red dotted line) and has a one sigma-width of $486$ km s$^{-1}$. Following \citetalias{Wetzell2022}, we identify outliers as clusters whose 68\% velocity dispersion lower limit is greater than the central velocity of the main population plus one standard deviation (black dot-dashed line). Outlier clusters chosen this way are shown in red with the main population clusters shown in blue.}
\label{fig:card_vel_resis}
\end{figure}

We quantitatively separate and define the outlier and main cluster populations following the methodology used in \citetalias{Wetzell2022}. To do this, we inspect the residuals of the cluster velocity dispersion compared to the $\sigma-\lambda$ relation found by \cite{Rozo2015b} using stacks of SDSS redMaPPer clusters. We fit a double Gaussian to the residuals as shown in Figure~\ref{fig:card_vel_resis} assuming that the populations have the same slope as the trend line from \cite{Rozo2015b}, but differing normalizations. 

We define outlier clusters as those whose lower limit on their velocity dispersion is more than one standard deviation away from the main velocity dispersion-richness relation (the primary peak in Figure \ref{fig:card_vel_resis}). Although the outlier population is not as well described by a Gaussian distribution as the main population is, the definition of outliers depends only on the fit to the main population. That is, the secondary component of the double Gaussian fit is not used for the definition for outliers, only to confirm their presence in the data. The median richnesses of the main and outlier populations are similar at 25.2 and 25.1 respectively.

As noted in \citetalias{Wetzell2022}, the normalization of the \cite{Rozo2015b} trend line lies above our main population despite the similar slopes. This is likely due to the fact that the stacked cluster velocity distributions used by \cite{Rozo2015b} would include both typical and outlier clusters. 

Compared to \citetalias{Wetzell2022}, our populations are fit fairly similarly. Table~\ref{tbl:pop_peaks} shows the peaks and widths of the main and outlier population for \citetalias{Wetzell2022}. 
The position of our main population is quite close to that found in \citetalias{Wetzell2022}, but we find a larger velocity width. The velocity center of our outlier distribution is somewhat lower, but not significantly. 
The larger width in both the main and outlier populations, $48$ and $148$ km s$^{-1}$ larger, respectively, indicates larger scatter in the velocity dispersions of the simulated clusters. However, the sample of observed clusters in \citetalias{Wetzell2022} is relatively small giving significant uncertainty and potentially leading to under sampling the scatter. 

The fraction of outlier clusters for the different samples discussed in this work are listed in Table \ref{tbl:outlier_frac}. Outliers make up approximately 35\% of the total Cardinal sample and are shown in red in Figure~\ref{fig:veldisp_richness}.  This is a significantly higher fraction than the 17\% outlier clusters found by \citet{Wetzell2022} in the DES Y3 data. However, as noted, the richness and redshift distributions of Cardinal clusters are not well-matched to the \citet{Wetzell2022} sample. In comparison, the Cardinal sample matched to \citetalias{Wetzell2022} has an outlier fraction of $\sim$20\%, similar to what was found in that paper.

\begin{table*}
    \centering
    \begin{tabular}{||c|c|c||} 
         \hline
         Population & Center Peak (km s$^{-1}$)  & $1\sigma$ Width (km s$^{-1}$)\\ [0.5ex] 
         \hline\hline
         Cardinal Main & -137 & 174  \\ 
         \hline
         Cardinal Outlier & 413 & 481  \\
         \hline
         \citetalias{Wetzell2022} Main & -131 & 126  \\
         \hline
         \citetalias{Wetzell2022} Outlier & 492 & 333  \\
         \hline
         Buzzard Main & -238 & 156  \\  
         \hline
         Buzzard Outlier & 354 & 489 \\
         \hline
    \end{tabular}
    \caption{Table of parameters of fitted Gaussians to the residuals between the mean value of \cite{Rozo2015b} and the main/outlier populations of Cardinal, \citetalias{Wetzell2022}, and Buzzard.}
    \label{tbl:pop_peaks}
\end{table*}

\begin{table*}
    \centering
    \begin{tabular}{||c|c||} 
         \hline
         Sample & Outlier Fraction \\ [0.5ex] 
         \hline\hline
         Cardinal & 35\%  \\ 
         \hline
         \citetalias{Wetzell2022} & 17\%   \\
         \hline
         Buzzard & 29\%  \\
         \hline
         Cardinal Matched to \citetalias{Wetzell2022}  & 20\% \\
         \hline
         Cardinal Matched to DES Y3 & 25\%  \\  
         \hline
         Cardinal for $0.2 < z < 0.65$ & 24\% \\
         \hline
         Cardinal Matched to DES Y3 for $0.2 < z < 0.65$ & 20\% \\
         \hline
    \end{tabular}
    \caption{Table of outlier fractions of Cardinal, \citetalias{Wetzell2022}, Buzzard, and our matched samples. Column 1 is the name of the sample and Column 2 is the outlier fraction.}
    \label{tbl:outlier_frac}
\end{table*}

Using the simulations, we can ask the question what fraction of outliers do we expect to be present in the full DES redMaPPer cluster catalogs. For this we consider the Cardinal sample matched to the DES Y3 volume-limited catalog. For this sample, we find an outlier fraction of approximately 25\%.  Considering only the redshift range $0.2 < z < 0.65$ utilized for DES cluster cosmology, the matched sample has an outlier fraction of 20\%. This is similar to the outlier fraction found both in \citetalias{Wetzell2022} and its matched Cardinal sample.

\subsection{Outliers and Relation to Redshift}
One of the primary difficulties with cluster selection from photometric data is contamination from nearby structure or galaxies in projection \citep{Sohn2018, Costanzi2019}. These projection effects can not only lead to a misidentification of cluster richness, but also a bias towards selecting clusters with nearby structure such as aligned filaments \citep{Costanzi2019, DES2020, Sunayama20}. If the outlier clusters are in regions with filaments or other groups/clusters along the line of sight, this could lead to higher velocity dispersions for a given richness.

\begin{figure*}
    \includegraphics[width = 1\linewidth]{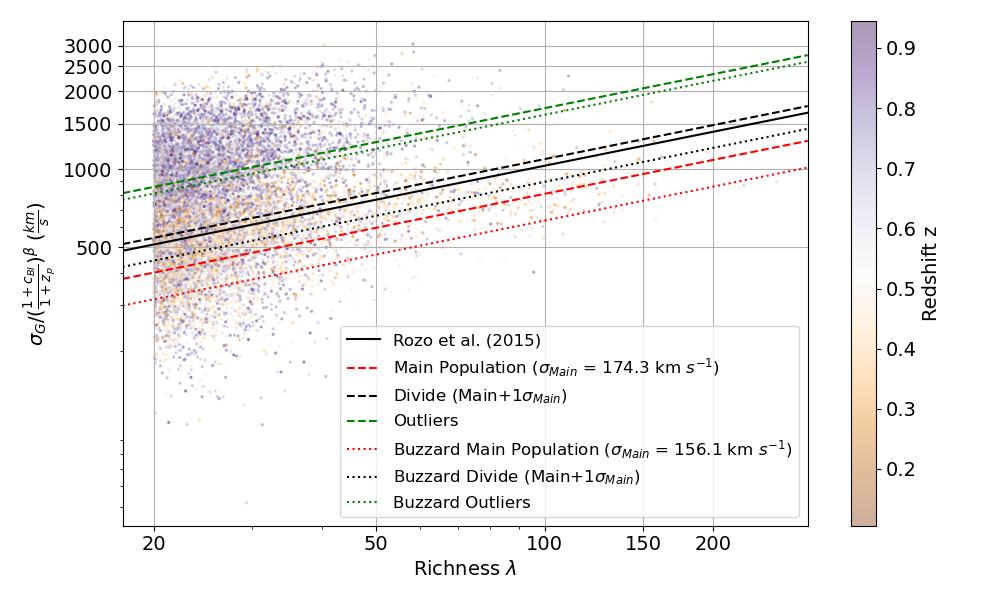}
    \caption{Same as Fig. 4, but with each cluster color-coded by their redshifts}
    \label{fig:veldisp_richness_zcolor}
\end{figure*}

\citetalias{Wetzell2022} found a higher frequency of velocity dispersion outliers at higher redshifts.  This is perhaps not surprising given to the difficulties of estimating photometric redshifts for high redshift galaxies and the widening of the red-sequence which can lead to increased projection effects. Looking at Figure~\ref{fig:veldisp_richness_zcolor}, which shows the velocity dispersion-richness relation color-coded by redshift, it is evident that the outlier population in the Cardinal simulations also has a higher average redshift than that of the main population. To test the discrepancy between the two distributions, we used a KS test. 
The KS test between the outlier redshift distribution and the main population redshift distribution gave a p-value essentially equal to zero, indicating a significant difference in these distributions.  

\begin{figure}
    \includegraphics[width = 1\linewidth]{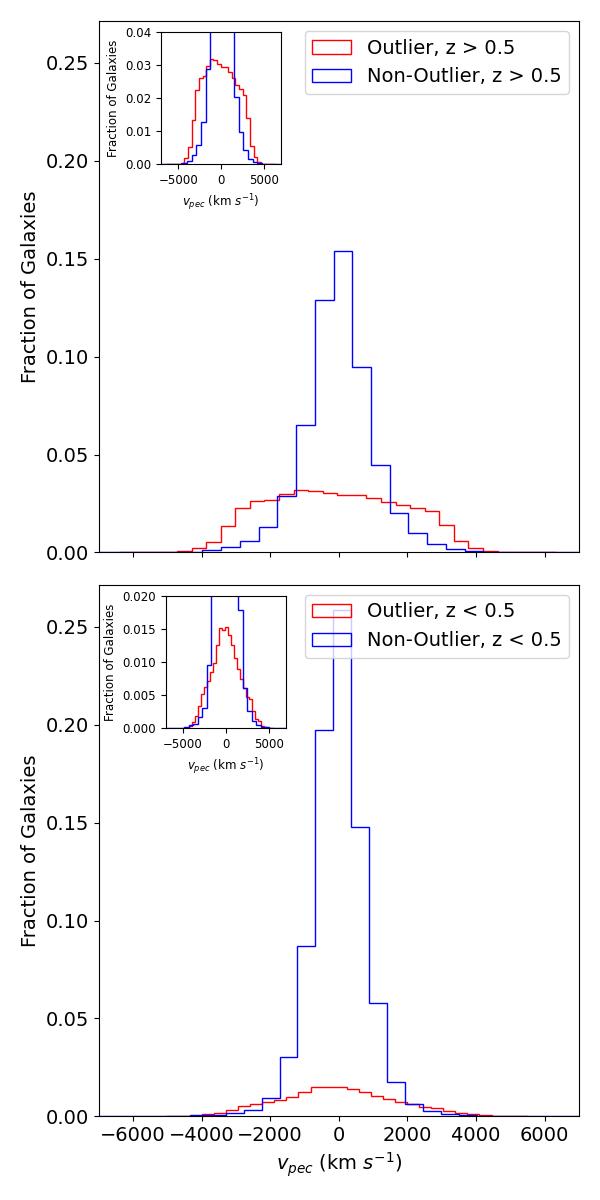}
    \caption{Histograms of the peculiar velocities of galaxies in clusters at high redshifts ($z > 0.5$; top panel) and at low redshifts ($z < 0.5$; bottom panel) for outlier clusters (red) and the main population (blue). Inset plots show the same histograms, but with a limited y-axis range.} 
    \label{fig:z_hist}
\end{figure}

A comparison of the stacked peculiar velocity distributions for clusters with both $z > 0.5$ and $z < 0.5$ in outlier and main populations,  is shown in Figure~\ref{fig:z_hist}. As can be seen in this figure, the outlier clusters have a much broader distribution of velocity offsets. Additionally, the outlier clusters can be seen to make up a larger fraction of the $z > 0.5$ clusters than in the $z < 0.5$ clusters. 

\subsection{Velocity Distributions}

\begin{figure*}
\includegraphics[width = 1\linewidth]{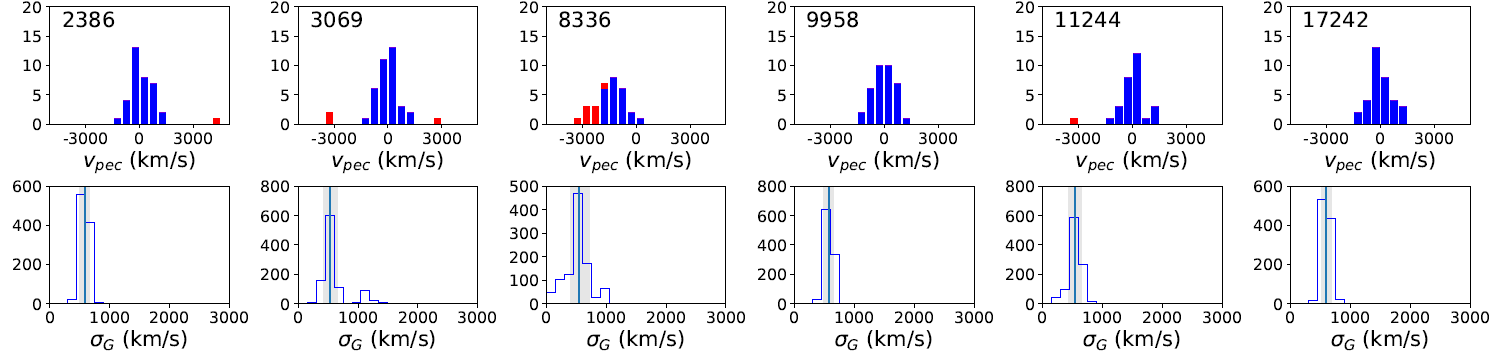}
\includegraphics[width = 1\linewidth]{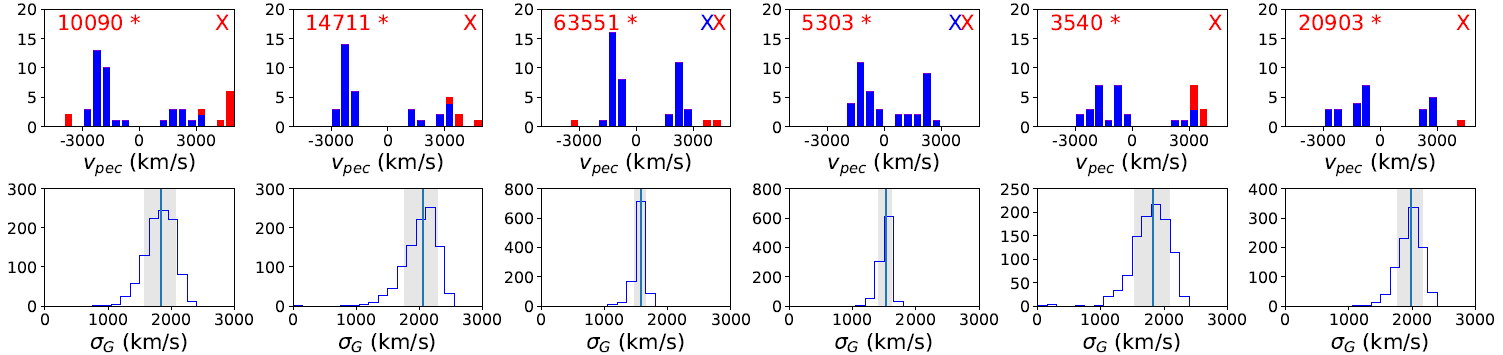}
\caption{A subsample of the gallery of cluster velocity distributions with corresponding $\sigma_{G}$ bootstraps. The top two rows correspond to clusters with $500\leq \sigma_{G} \leq 600$ and $20\leq \lambda \leq 30$ while the bottom two rows correspond to clusters with $1500 \leq \sigma_{G}$ and $20\leq \lambda \leq 30$. The MEM\_MATCH\_ID is shown in the top left of each subplot with outliers being marked in red with an asterisk. The red X in the top right denotes if a cluster was found to be non-Gaussian via the Anderson-Darling (AD) goodness-of-fit test while the blue X denotes if the cluster has significant substructure found via the Dressler-Shectman (DS) test. For the peculiar velocity graphs ($v_{pec}$), member galaxies are shown in blue and interloping galaxies are shown in red. For the $\sigma_{G}$ bootstrap graphs, the black line shows our reported $\sigma_{G}$ and the gray bar shows our $\sigma_{G}$ 68$\%$ confidence interval.}
\label{fig:veldist}
\end{figure*}

To further investigate the nature of the outlier clusters, we look at the velocity distributions of individual clusters in our sample to determine if they differ between the main population and the outlier population. Figure~\ref{fig:veldist} shows galleries of the velocity offset distributions along with histograms of the bootstrap resampling of the velocity dispersions for clusters in two subsamples. The first gallery contains clusters with velocity dispersions between $500$ km s$^{-1}$ and $600$ km s$^{-1}$ and richnesses between 20 and 30. The second gallery contains clusters with velocity dispersions $\geq 1500$ km s$^{-1}$ and richnesses between 20 and 30. These ranges are meant to pick out clusters with similar richness in the main population and the outlier population, respectively. Figure~\ref{fig:veldist} shows a portion of each gallery with the top two rows showing clusters in the main population and the bottom two rows showing the outlier population; the full gallaries for clusters in these bins are shown in Appendix A. A relaxed cluster should theoretically have a Gaussian velocity distribution, however, the velocity distributions of the outlier population look to be more non-Gaussian than the main population, often having multiple peaks or being spread out. 

Using the Anderson-Darling test, we find that approximately 80\% (733/980) of clusters with $500\leq \sigma_{G} \leq 600$ and $20\leq \lambda \leq 30$ are considered Gaussian by the AD Test. In comparison, only $\sim2$\% (31/1270) of clusters with $\sigma_{G} \geq 1500$ and $20\leq \lambda \leq 30$ are Gaussian by the AD Test. Given that interlopers are mostly removed by initial cuts, it seems likely that the non-Gaussianity of the outlier clusters is due to structure along the line-of-sight and projection effects. 

We also investigate the presence of substructure using the DS test. Given the high level of non-Gaussianity of the outlier population compared to the main population, we expect that outlier clusters would be more likely to contain some substructure. We find that approximately 14\% (131/920) of clusters with $500\leq \sigma_{G} \leq 600$ and $20\leq \lambda \leq 30$ are considered to have significant substructure. In comparison, the outlier population shows approximately 24\% (311/1270) of clusters with $\sigma_{G} \geq 1500$ and $20\leq \lambda \leq 30$ to have significant substructure. While the outlier population does have a significantly higher fraction of clusters with substructure, it is not a majority of outlier clusters. This could be due to our conservative definition of clusters with substructure; as previously mentioned the DS test should be taken as a lower limit on the percentage of clusters with substructure. However, this result might also indicate that substructure only partially explains the overestimation of velocity dispersion in the outlier clusters.  
 
\subsection{Relation to Halo Mass}
An important question is whether or not the redMaPPer richness and/or velocity dispersion are good tracers of halo mass for the outlier clusters.  Here we examine these properties compared to halo mass using $M_{500c}$ from the Cardinal simulations with halos matched to clusters as described in Section~\ref{subsec:halo_match}. 
Additionally, we compare the halo mass relations between the main population and outlier clusters. 

\begin{figure*}
\includegraphics[width = 1\linewidth]{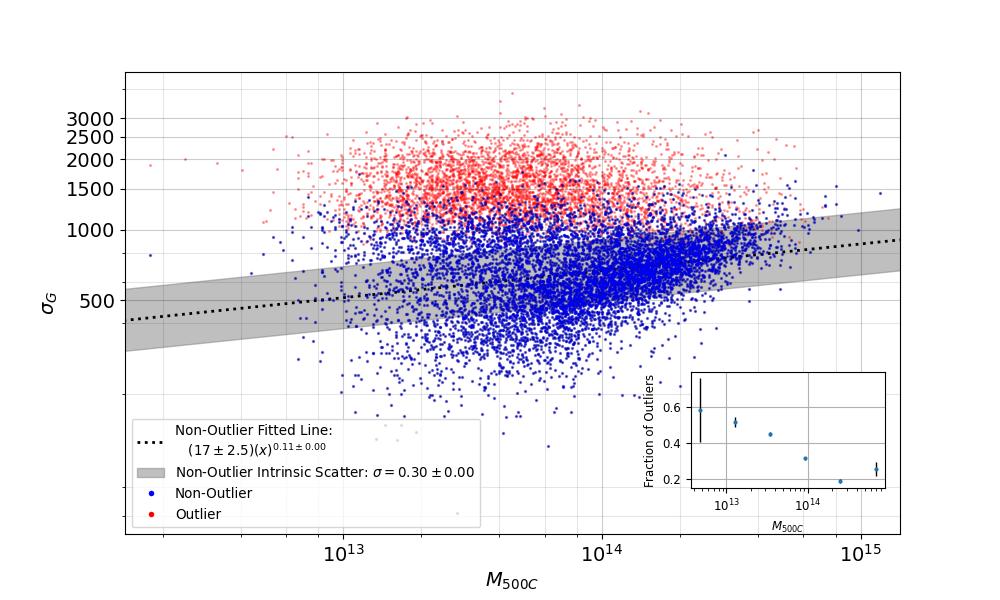}
\caption{The relation between cluster velocity dispersion, $\sigma_{G}$, and halo mass, $M_{500c}$. The main population is shown in blue and the outlier population is shown in red. The dotted black line is the $\sigma_{G}$-$M_{500c}$ relation fitted to the main population with the shading showing the $1\sigma$ scatter. The inset graph shows the mass distribution of the outlier fraction with error bars coming from propagating the error of outlier cluster counts over all cluster counts for each bin.}
\label{fig:sigm}
\end{figure*}

By definition, the outlier clusters have high velocity dispersion for their richness, and Figure~\ref{fig:sigm}, which shows the relation between the velocity dispersion and halo mass, indicates that these clusters also have velocity dispersions that are high compared to their halo masses.  Here we use the $M_{500c}$ halo mass, the mass inside a radius within which the density is 500 times the critical density of the universe at the cluster redshift. In this figure, the main population is shown in blue while the outlier population is shown in red. 

Since we expect velocity dispersion to increase with mass, one could expect a population of high velocity dispersion clusters to have higher masses than those of low velocity dispersion clusters. However, both the main and outlier populations have roughly the same range of halo masses with the outliers simply offset to higher $\sigma_G$. The outlier fraction is also higher for lower masses, and the outlier clusters do not show the expected correlation between $\sigma_G$ and mass. This further indicates that the outlier clusters are the result of overestimating velocity dispersions due to line-of-sight structure. The main population instead does show a generally tight relationship between $\sigma_G$ and halo mass with a fitted scatter of $\sigma \sim 0.30$, though with residual extra scatter at lower masses.  Neither Figure~\ref{fig:sigm} nor the $\sigma_G - \lambda$ relation in Figure 2 show a sharp division between the outlier and main cluster populations, and our outlier cut appears to be somewhat conservative in the sense that there are intermediate clusters with velocity dispersions relatively high for their mass that do not meet our outlier cut.

\begin{figure*}
\includegraphics[width = 1\linewidth]{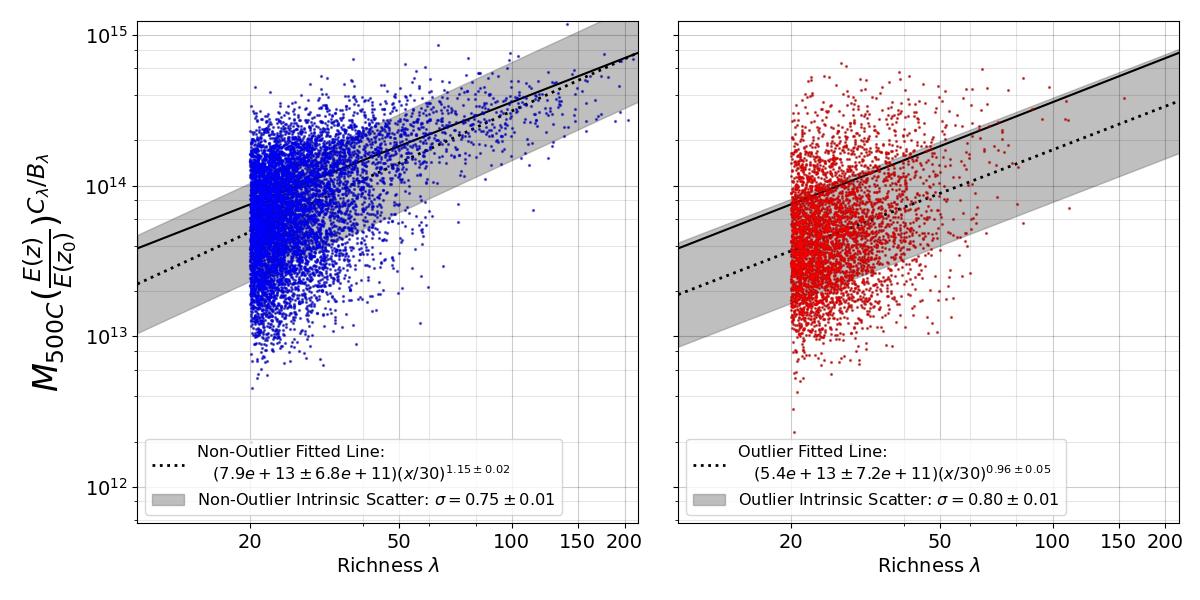}
\caption{The relation between cluster richness, $\lambda$, and halo mass, $M_{500c}$ for the main and outlier populations. The main population is shown on the left in blue while the outliers are shown on the right in red. The black line is the $\lambda$-$M_{500c}$ trendline from \cite{Grandis21}. The dotted black line is the $\lambda$-$M_{500c}$ relation fitted to each population at a pivot richness of 30 with the shading showing the $1\sigma$ scatter. The redshift dependence is accounted for by weighting the cluster halo mass by E(z) ($z_{0} = 0.6, C_{\lambda} = 0.69$) following \cite{Grandis21}.} 
\label{fig:lamm}
\end{figure*}

Figure~\ref{fig:lamm} shows the relation between halo mass and the cluster richness, $\lambda$. The left graph shows the relation for main population clusters and the right graph shows the relation for outlier clusters. The main population shows the expected correlation between richness and mass with a scatter of $\sigma = 0.75$, though with higher scatter at low richness. The high scatter at low richness could indicate that selection biases are particularly strong around $20\leq\lambda\leq40$.
In fact, based on the SPT confirmation fraction, \citet{Grandis21} find that the scatter or contamination fraction of the redMaPPer DES Y1 sample grows with decreasing richness below $\lambda \sim 40$. 

The halo mass relation of the outlier clusters shows a similar overall range to the main population. However, the outlier clusters have a weaker correlation of richness with mass then seen in main population and possess overall higher scatter with a value of $\sigma = 0.80$. The outliers are also offset to somewhat lower halo masses, which can be seen in the offset in normalization in Figure~\ref{fig:lamm}. The average halo mass for the main population clusters with $\lambda \leq 50$ is approximately $M_{500c}=9.1\times10^{13}\;h^{-1}_{70}\: M_{\odot}$ while the average halo mass for outlier clusters with $\lambda \leq 50$ is approximately $M_{500c}=6.7\times10^{13}\;h^{-1}_{70}\: M_{\odot}$.  While the large, low-richness scatter on the richness-mass relation is not entirely due to the outlier clusters, it is clear that these clusters add to the scatter/contamination of redMaPPer-selected clusters.

To further explore the validity of redMaPPer richness, we can use the halo matching to determine a ``true" richness for clusters by counting the number of galaxies that live within a given matched halo as described in Section~\ref{subsec:halo_match}. It is worth noting here that $R_\lambda$, the radius employed by redMaPPer to identify potential member galaxies, is not the same as $R_{200}$, the cut off for considering a galaxy to be part of a halo; in general $R_{200} > R_\lambda$ for high-mass clusters, and $R_{200} < R_\lambda$ for low mass clusters. We then compare this ``true" halo richness to the redMaPPer richness and to halo mass. 

\begin{figure*}
\includegraphics[width = 1\linewidth]{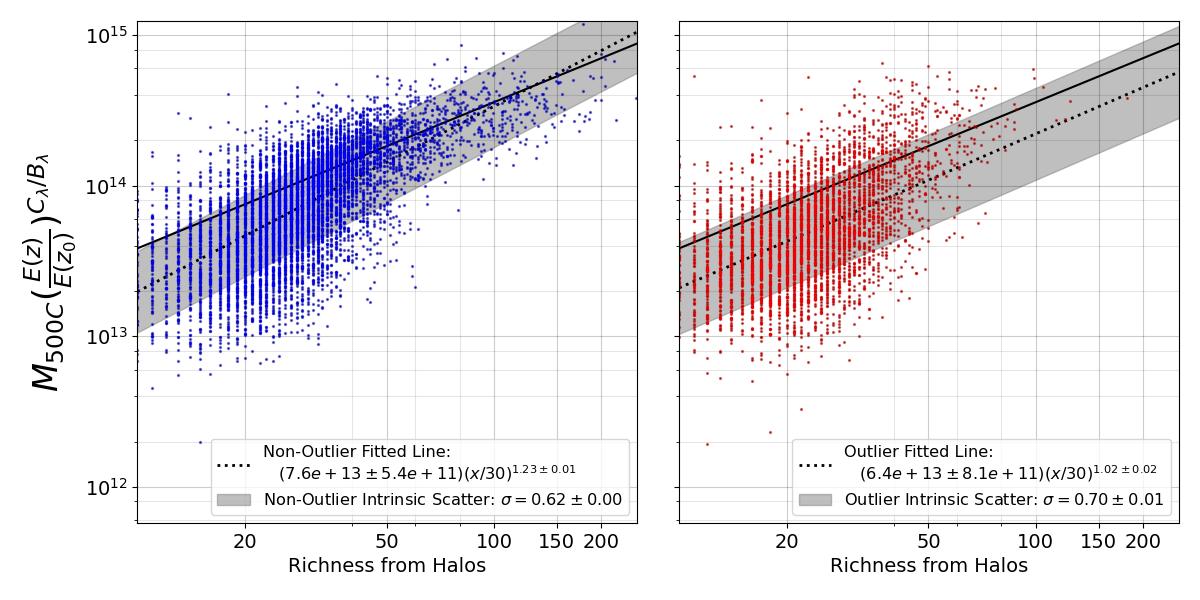}
\caption{The relation between the halo richness, determined by counting the number of redMaPPer identified galaxies within a given halo, and the halo mass, $M_{500c}$ for the main and outlier population. The main population is shown on the left in blue while the outliers are shown on the right in red. The black line is the $\lambda$-$M_{500c}$ trendline from \cite{Grandis21}. The dotted black line is the $\lambda$-$M_{500c}$ relation fitted to the population with the shading showing the $1\sigma$ scatter. The redshift dependence is accounted for weighting the cluster halo mass by cluster redshift ($z_{0} = 0.6, C_{\lambda} = 0.69$) following \cite{Grandis21}.}
\label{fig:truelamm}
\end{figure*}

Figure~\ref{fig:truelamm} shows the relation between the halo mass and true richness. The left graph shows the relation for main population clusters and the right graph shows the relation for outlier clusters. Both populations show some of the expected correlation between richness and mass.  In particular, using the true richness there is a stronger correlation with halo mass for the outlier clusters, though significant scatter remains. 
The scatter in both populations is lower than in the redMaPPer richness-mass relation with the main and outlier populations having scatters of $\sigma = 0.62$ and $\sigma = 0.70$, respectively. However, even using true richness there is still prominent scatter at low richnesses. 

\begin{figure*}
    \includegraphics[width = 1\linewidth]{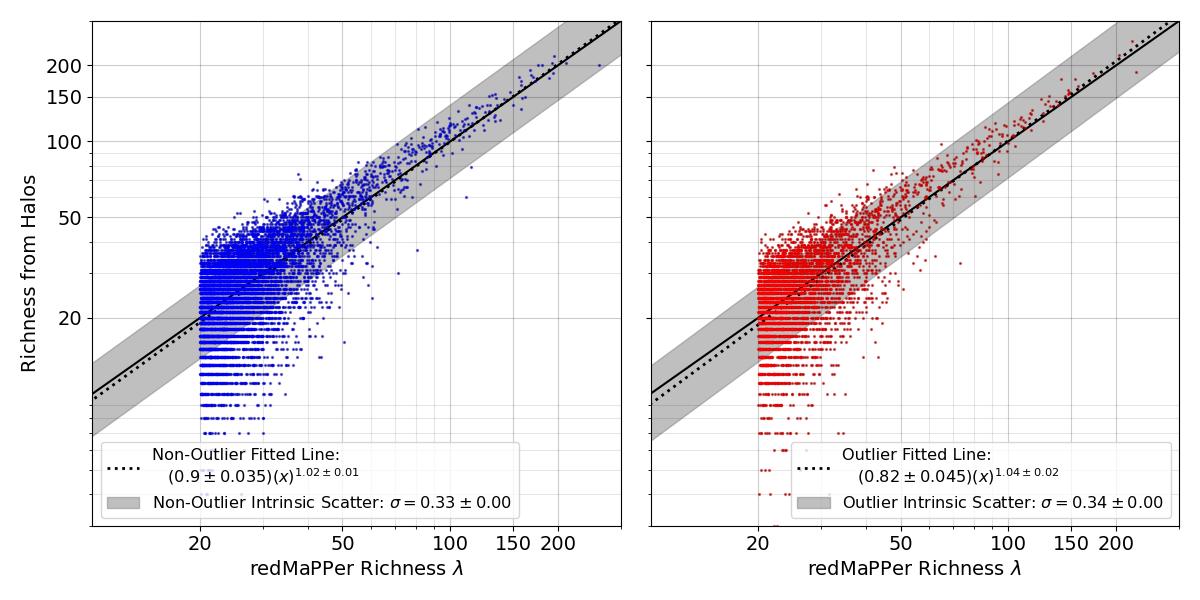}
    \caption{The relation between the redMaPPer richness for the main and outlier populations and the ``true" richness determined by counting the galaxies within a given halo. The main population is shown on the left in blue while the outliers are shown on the right in red. The black line represents a one-to-one linear relation. The dotted black line is the fitted relation to each population with the shading showing the $1\sigma$ scatter.}
    \label{fig:truelam}
\end{figure*}

\begin{figure}
    \includegraphics[width = 1\linewidth]{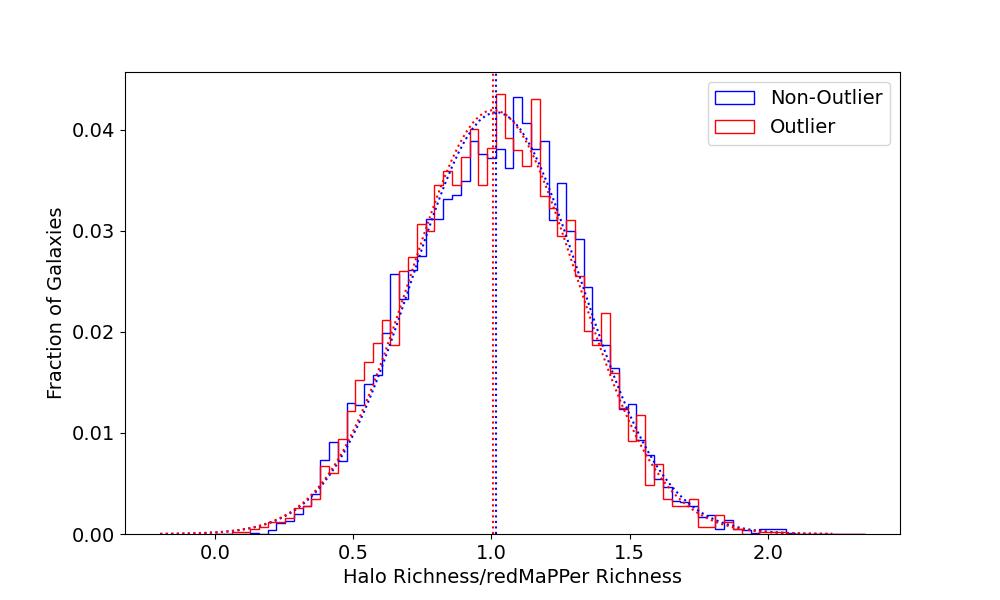}
    \caption{The normalized distributions of the ratio between the ``true" richness determined from halo data and the redMaPPer richness for the main (blue) and outlier populations (red). The dotted lines represent the Gaussians fitted to the population with the corresponding color with the vertical dotted lines representing the center of each Gaussian. 
    }
    \label{fig:truelamratio}
\end{figure}

Figure~\ref{fig:truelam} shows the relation between the halo richness and the redMaPPer richness, $\lambda$. Both the main and outlier clusters show a strong correlation between the redMaPPer estimated richness and the halo richness, but again with large scatter at low richness. Interestingly, for the main cluster population, the redMaPPer richness appears to be an underestimate of the true richness potentially indicating that the membership probabilities are too low, though the redMaPPer richness does not need to be the true richness if it correlates well with mass. 

To further investigate, Figure~\ref{fig:truelamratio} shows the distributions of true richness divided by redMaPPer richness for the main and outlier populations. These distributions show that the underestimation of the true richness is very minor given that both are centered very close to, but slightly above one. The main population is centered at 1.02 and the outlier population is centered at 1.01. Both populations look similar and have similar scatter, indicating that redMaPPer richness is an equally good indicator of true richness for both populations. 

\section{Comparison to Buzzard}
As an indication of the potential dependence of the outlier cluster population on the specific prescription used in creating the mock catalogs, 
we compare our results for Cardinal to the previous version of the DES mocks known as Buzzard \citep{derose2019}. We used Buzzard version 1.9.8 and a final sample of 7413 clusters. Cardinal contains improved subhalo abundance matching and an updated color assignment model compared to Buzzard \citep{To2023}. Buzzard also significantly underestimates the number of galaxy clusters at a given richness for redMaPPer clusters when compared to DES Y1 data \citep{derose2019, Abbott2019}. 
For these reasons, our analysis of Buzzard is not as in depth as that of Cardinal. However, it is still a good test of whether the prescription for assigning galaxy properties to mock catalogs impacts  the presence of outlier clusters. 


\begin{figure*}
    \includegraphics[width = 1\linewidth]{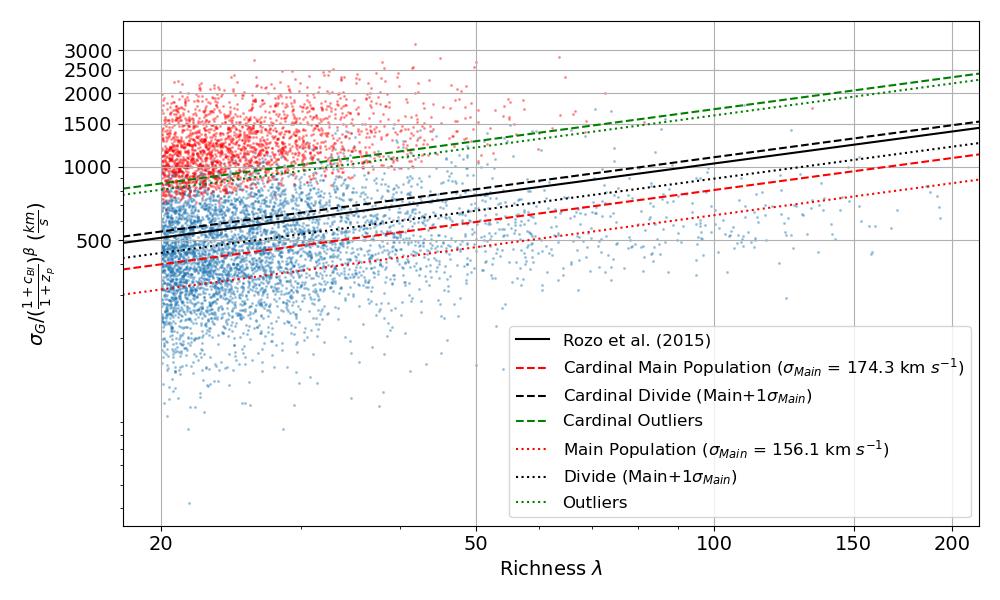}
    \caption{Similar to Figure~\ref{fig:veldisp_richness}, but for the Buzzard catalog. 
    }
    \label{fig:buzz_sig_lam}
\end{figure*}

Figure~\ref{fig:buzz_sig_lam} shows the velocity dispersion-richness relation for the Buzzard mocks. As with the Cardinal simulations, the relation follows a power law with a slope of $\sim0.44$ with a bi-modal population. Similarly, there also appears to be a population with high velocity dispersions for a given richness. 

\begin{figure}
    \includegraphics[width = 1\linewidth]{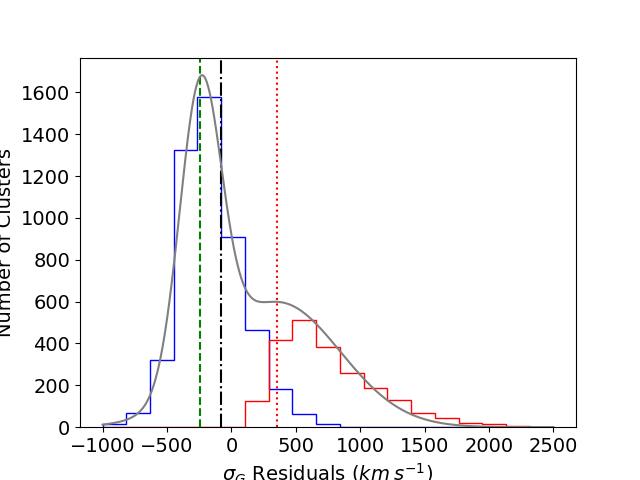}
    \caption{Similar to Figure~\ref{fig:card_vel_resis}, but for the Buzzard catalog. 
    }
    \label{fig:buzz_cut}
\end{figure}

To investigate this relation, we again fit a double Gaussian to the residuals of the cluster velocity dispersions compared to the \citet{Rozo2015b} relation as seen in Figure~\ref{fig:buzz_cut} with the fit parameters listed in Table \ref{tbl:pop_peaks}. For comparison, we show the Buzzard fit lines in Figures \ref{fig:veldisp_richness} and \ref{fig:veldisp_richness_zcolor}, and the Cardinal fit lines in Figure \ref{fig:buzz_sig_lam}. 

In Buzzard, the outlier population makes up around 29\% of the clusters. This is somewhat lower than the fraction in Cardinal, but still higher than the fraction found by \citetalias{Wetzell2022}. Using matched samples analogous to those created for Cardinal, the outlier population makes up around 14\% in the \citetalias{Wetzell2022} matched sample and 29\% in the DES matched sample. These values differ somewhat from those in Cardinal, being lower and higher, respectively. Some discrepancy in the outlier fractions is expected given the different prescriptions for subhalos and galaxy colors. 
For example, it is known that the Buzzard simulations contain too few red satellite galaxies in clusters \citep{derose2019}. However, even with these known discrepancies, the values do not differ greatly. 

The fits to the $\sigma_g-\lambda$ relation in Buzzard are significantly different from \citetalias{Wetzell2022} much more so than in Cardinal. Both the main and outlier peaks are centered at much lower velocity dispersion values than the peaks of Cardinal and \citetalias{Wetzell2022}. However, the sigma-width of both populations is similar to those in Cardinal. The main population's sigma-width is also close to the one found in \citetalias{Wetzell2022}; however, as with Cardinal, the outlier population's one sigma-width is higher. 



\begin{figure*}
    \includegraphics[width=1\linewidth]{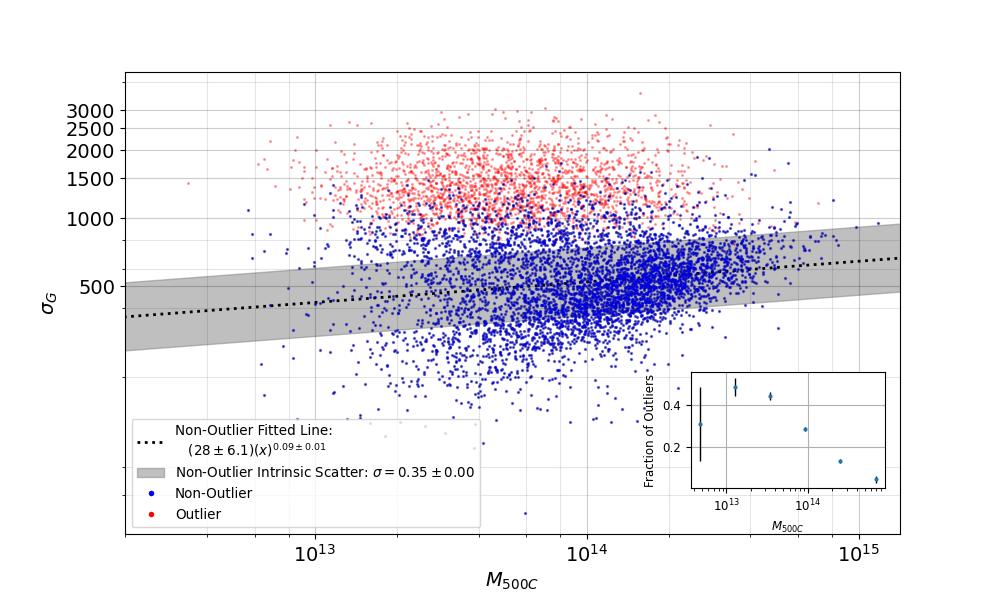}
    \caption{Similar to Figure~\ref{fig:sigm}, but for the Buzzard catalog.
    }
    \label{fig:buzz_sigm}
\end{figure*}

\begin{figure*}
    \includegraphics[width=1\linewidth]{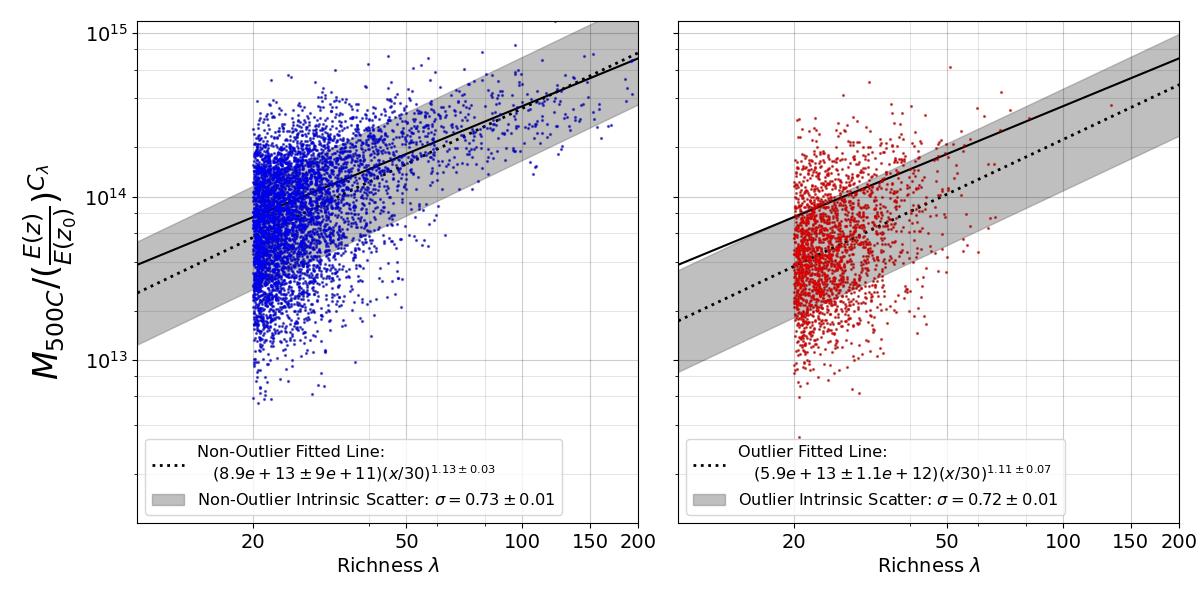}
    \caption{Similar to Figure~\ref{fig:lamm}, but for the Buzzard catalog.}
    \label{fig:buzz_lamm}    
\end{figure*}

Figure~\ref{fig:buzz_sigm} and Figure~\ref{fig:buzz_lamm} show the velocity dispersion to halo mass ($M_{500c}$) relation and the halo mass to cluster richness relation, respectively for the Buzzard catalogs.  These figures show similar trends to those seen in Cardinal.  The outlier clusters show no strong correlation of velocity dispersion with halo mass, and for the same richness the outliers tend to have lower halo mass. While the main population does show correlations between velocity dispersion, richness, and mass, the scatter is high for lower richnesses or halo masses.

\section{Conclusion}
In this paper, we investigate aspects of the redMaPPer cluster finder selection using the Cardinal mock galaxy catalogs \citep{To2023}.  In particular, we look at the peculiar velocity distributions and velocity dispersions of redMaPPer member galaxies as an indication of the line-of-sight structure and the relationship to richness, redshift, and host halo mass.  We also compare these results to the Buzzard simulations.

Looking at the velocity dispersion of redMaPPer member galaxies, we find a large population of clusters whose velocity dispersions are high for their richnesses.  A similar population was found in the DES Y3 data by \citetalias{Wetzell2022}. 
When constructing samples matched in richness and redshift distribution, we find a similar fraction of outliers in the simulations and in data.  Our matched results indicated that we expect 25\% of DES Y3 redMaPPer clusters and 20\% in the redshift range used for cosmology to be velocity dispersion outliers.  
The analogous Buzzard to DES matched sample would imply a somewhat larger fraction of outliers, but the fractions are largely similar given differences in the prescriptions for galaxies between the two simulations.

Another 3\% of Cardinal clusters had too few members to find a velocity dispersion following our initial cut on peculiar velocity to remove interlopers, meaning that a significant fraction of galaxies which are putative cluster members in these clusters were the result of projection.  In both the Cardinal and Buzzard simulations, the scatter in velocity dispersions is somewhat larger than seen in the data, though the sample size in \citetalias{Wetzell2022} was relatively small.

The peculiar velocity distributions of the velocity dispersion outliers are non-Gaussian and have a higher prevalence of substructure.  Comparing clusters with richnesses between $20-30$ that are outliers and to those with similar richness in the main population, we find that 98\% of the outliers have non-Gaussian velocity distributions using the Anderson-Darling test compared to only 20\% of the main population. A Dressler-Shectman test shows that a higher fraction of outliers contain significant substructure in their velocity distributions, indicating multiple groups along the line of sight.  Specifically, 24\% of outliers have significant substructure compared to 14\% of the main population.  We note that our substructure criterion based on a Dressler-Shectman test significance greater than 95\% is conservative, so a higher fraction of clusters may have substructure.
The high non-Gaussianity and slightly higher presence of substructure in the outlier clusters indicates the selection by redMaPPer of complex sight lines with significant line-of-sight structure.

Examination of the relations of richness and velocity dispersion to halo mass further indicate that the presence of the outliers is due to line-of-sight structure. The main population shows the expected correlation between both richness and velocity dispersion with mass, though with large scatter in halo mass at low richness ($\lambda \lesssim 50$).  The outlier clusters, however, show little to no trend in observed richness and halo mass with larger scatter than in the main population.
While scatter is present in the main population, it is clear that the outlier clusters contribute to the scatter/contamination of redMaPPer-selected clusters.  The outliers also have high velocity dispersions for their mass and little to no correlation between velocity dispersion and mass.  The Buzzard simulations show very similar trends with halo mass as seen in Cardinal.

We explore the validity of redMaPPer richness by deriving a ``true" richness as described in Section~\ref{subsec:halo_match}. 
Using this true richness, the scatter in the richness-mass relation is reduced for both typical and outlier clusters with both populations having their scatter decrease by $\sim 0.10$. Additionally, the true richnesses of outlier clusters show a better correlation with halo mass, albeit with large scatter.

Our results indicate that the presence of outlier clusters is most likely due to the overestimation of velocity dispersions due to line-of-sight structure and projection effects. Additionally, projection effects and line-of-sight structure contribute to cluster selection and richness estimation in redMaPPer, notably at lower richnesses and at higher redshifts. 

Our results give an indication of the frequency of line-of-sight selection effects and their contribution to richness scatter, which can be used to help calibrate redMaPPer selection. Going forward further work with simulations could explicitly connect velocity dispersion outliers to the line-of-sight halo distribution and the related bias in weak lensing shear.  Simulations could also potentially be used to improve richness and membership probability estimates.

\section*{Acknowledgements}
Funding for the DES Projects has been provided by the U.S. Department of Energy, the U.S. National Science Foundation, the Ministry of Science and Education of Spain, 
the Science and Technology Facilities Council of the United Kingdom, the Higher Education Funding Council for England, the National Center for Supercomputing 
Applications at the University of Illinois at Urbana-Champaign, the Kavli Institute of Cosmological Physics at the University of Chicago, 
the Center for Cosmology and Astro-Particle Physics at the Ohio State University,
the Mitchell Institute for Fundamental Physics and Astronomy at Texas A\&M University, Financiadora de Estudos e Projetos, 
Funda{\c c}{\~a}o Carlos Chagas Filho de Amparo {\`a} Pesquisa do Estado do Rio de Janeiro, Conselho Nacional de Desenvolvimento Cient{\'i}fico e Tecnol{\'o}gico and 
the Minist{\'e}rio da Ci{\^e}ncia, Tecnologia e Inova{\c c}{\~a}o, the Deutsche Forschungsgemeinschaft and the Collaborating Institutions in the Dark Energy Survey. 

The Collaborating Institutions are Argonne National Laboratory, the University of California at Santa Cruz, the University of Cambridge, Centro de Investigaciones Energ{\'e}ticas, 
Medioambientales y Tecnol{\'o}gicas-Madrid, the University of Chicago, University College London, the DES-Brazil Consortium, the University of Edinburgh, 
the Eidgen{\"o}ssische Technische Hochschule (ETH) Z{\"u}rich, 
Fermi National Accelerator Laboratory, the University of Illinois at Urbana-Champaign, the Institut de Ci{\`e}ncies de l'Espai (IEEC/CSIC), 
the Institut de F{\'i}sica d'Altes Energies, Lawrence Berkeley National Laboratory, the Ludwig-Maximilians Universit{\"a}t M{\"u}nchen and the associated Excellence Cluster Universe, 
the University of Michigan, NSF NOIRLab, the University of Nottingham, The Ohio State University, the University of Pennsylvania, the University of Portsmouth, 
SLAC National Accelerator Laboratory, Stanford University, the University of Sussex, Texas A\&M University, and the OzDES Membership Consortium.

Based in part on observations at NSF Cerro Tololo Inter-American Observatory at NSF NOIRLab (NOIRLab Prop. ID 2012B-0001; PI: J. Frieman), which is managed by the Association of Universities for Research in Astronomy (AURA) under a cooperative agreement with the National Science Foundation.

The DES data management system is supported by the National Science Foundation under Grant Numbers AST-1138766 and AST-1536171.
The DES participants from Spanish institutions are partially supported by MICINN under grants PID2021-123012, PID2021-128989 PID2022-141079, SEV-2016-0588, CEX2020-001058-M and CEX2020-001007-S, some of which include ERDF funds from the European Union. IFAE is partially funded by the CERCA program of the Generalitat de Catalunya.

We  acknowledge support from the Brazilian Instituto Nacional de Ci\^encia
e Tecnologia (INCT) do e-Universo (CNPq grant 465376/2014-2).

This manuscript has been authored by Fermi Research Alliance, LLC under Contract No. DE-AC02-07CH11359 with the U.S. Department of Energy, Office of Science, Office of High Energy Physics.

\bibliography{refs}{}
\bibliographystyle{aasjournal}

\appendix
\setcounter{figure}{0}
\renewcommand{\thefigure}{A\arabic{figure}}
\section{Full Bootstrap and Velocity Distribution Catalogs}
In this appendix, we present the galleries of clusters' velocity distribution histograms with the corresponding $\sigma_{G}$ bootstraps used in Figure~\ref{fig:veldist}. The first gallery corresponds to clusters with $500\leq \sigma_{G} \leq 600$ and $20\leq \lambda \leq 30$ (Figure~\ref{fig:veldistnonoutlier}). The second gallery corresponds to clusters with $1500 \leq \sigma_{G}$ and $20\leq \lambda \leq 30$ (Figure~\ref{fig:veldistoutlier}). The MEM\_MATCH\_ID is shown in the top left of each subplot with outliers being marked in red with an asterisk. The red X in the top right denotes if a cluster was found to be non-Gaussian via the Anderson-Darling (AD) goodness-of-fit test while the blue X denotes if the cluster has significant substructure found via the Dressler-Shectman (DS) test. For the peculiar velocity graphs ($v_{pec}$), member galaxies are shown in blue with interloping galaxies being shown in red. For the $\sigma_{G}$ bootstrap graphs, the black line shows our reported $\sigma_{G}$ and the gray bar shows our $\sigma_{G}$ confidence interval. For the sake of brevity, only the first three pages of each gallery are shown in this appendix.  

\begin{figure}
      \centering
      \caption{The gallery of clusters that have $500\leq \sigma_{G} \leq 600$ and $20\leq \lambda \leq 30$ with corresponding $\sigma_{G}$ bootstraps. The MEM\_MATCH\_ID is shown in the top left of each subplot with outliers being marked in red with an asterisk. The red X in the top right denotes if a cluster was found to be non-Gaussian via the Anderson-Darling (AD) goodness-of-fit test while the blue X denotes if the cluster has significant substructure found via the Dressler-Shectman (DS) test. For the peculiar velocity graphs ($v_{pec}$), member galaxies are shown in blue with interloping galaxies being shown in red. For the $\sigma_{G}$ bootstrap graphs, the black line shows our reported $\sigma_{G}$ and the gray bar shows our $\sigma_{G}$ confidence interval.}
      \includegraphics[scale = 0.45]{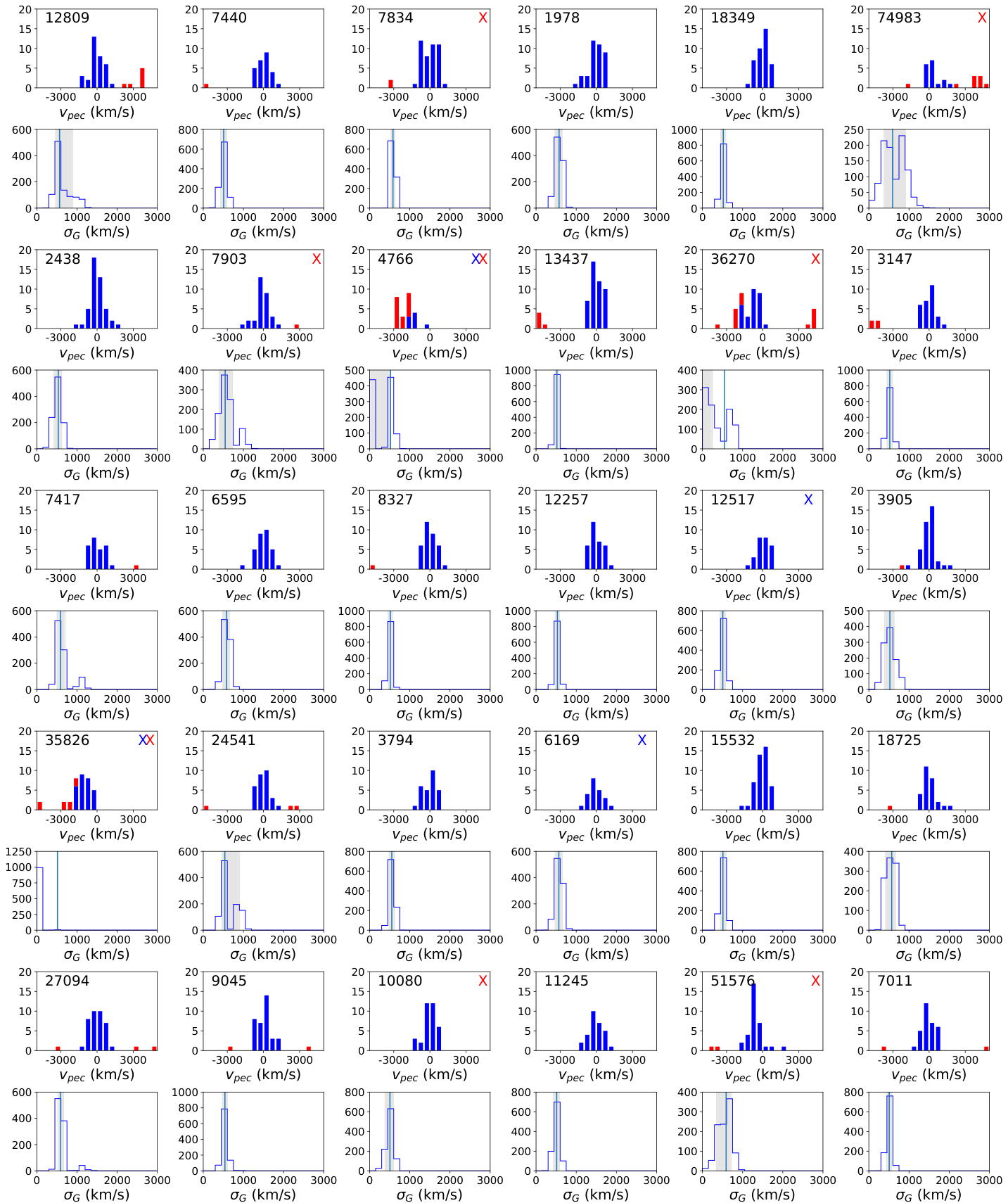}
      \label{fig:veldistnonoutlier}
\end{figure}

\begin{figure}
    \ContinuedFloat
    \centering
    \subfloat{\includegraphics[scale = 0.45]{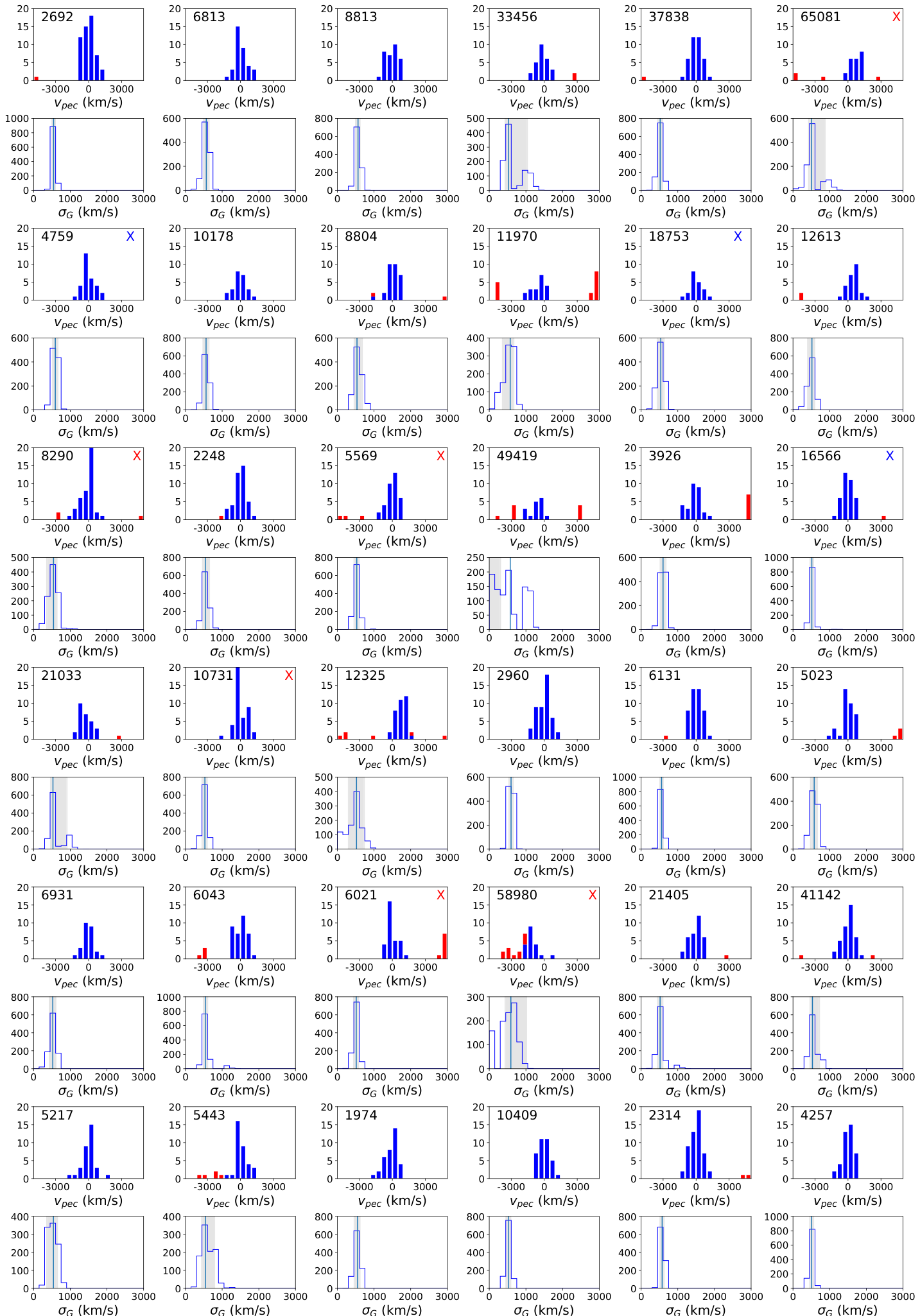}}
\end{figure}

\begin{figure}
    \ContinuedFloat
    \centering
    \subfloat{\includegraphics[scale = 0.45]{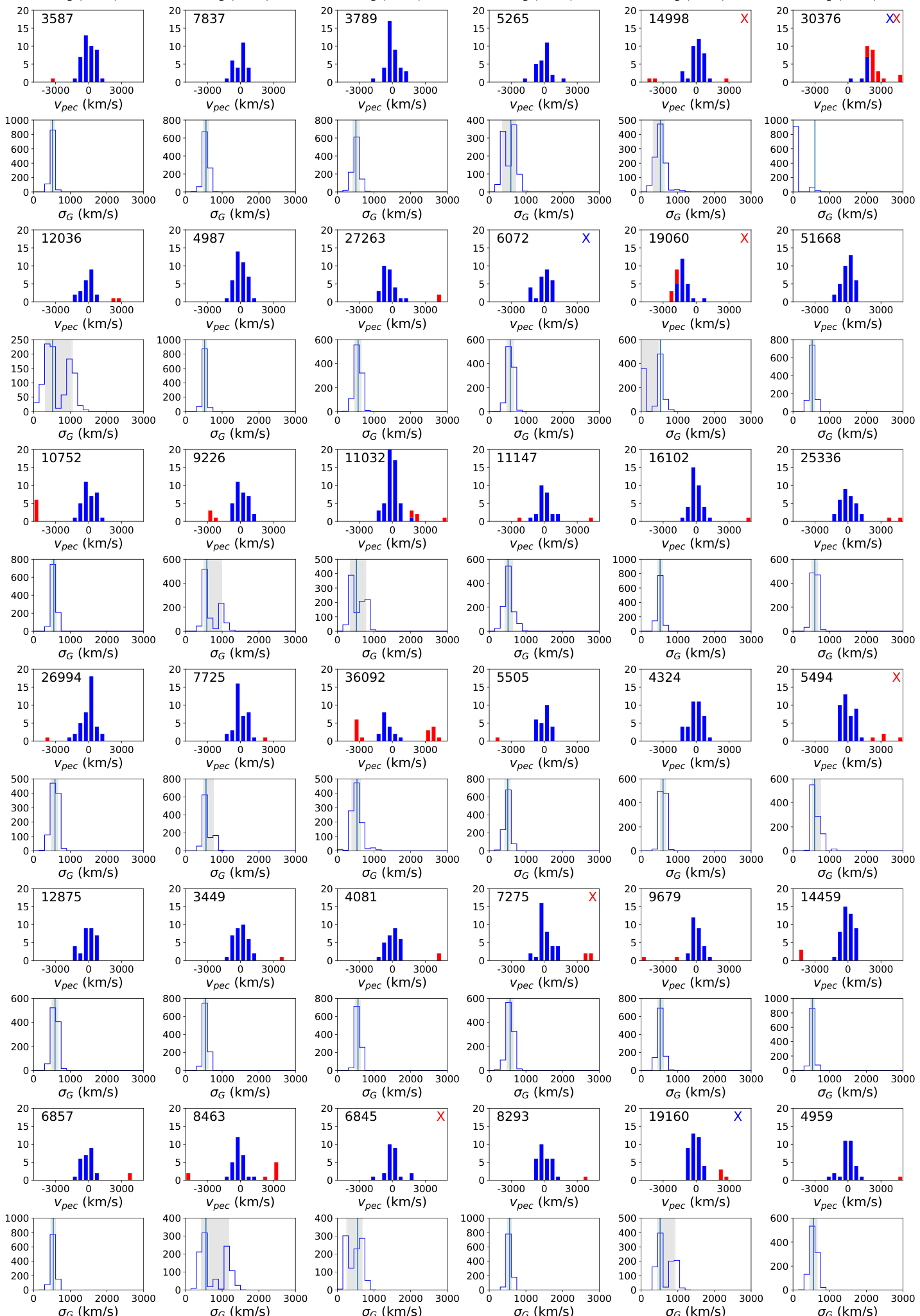}}
\end{figure}

\setcounter{figure}{1}

\begin{figure}
      \centering
      \caption{The gallery of clusters that have $1500 \leq \sigma_{G}$ and $20\leq \lambda \leq 30$ with corresponding $\sigma_{G}$ bootstraps. The MEM\_MATCH\_ID is shown in the top left of each subplot with outliers being marked in red with an asterisk. The red X in the top right denotes if a cluster was found to be non-Gaussian via the Anderson-Darling (AD) goodness-of-fit test while the blue X denotes if the cluster has significant substructure found via the Dressler-Shectman (DS) test. For the peculiar velocity graphs ($v_{pec}$), member galaxies are shown in blue with interloping galaxies being shown in red. For the $\sigma_{G}$ bootstrap graphs, the black line shows our reported $\sigma_{G}$ and the gray bar shows our $\sigma_{G}$ confidence interval.}
      \subfloat{\includegraphics[scale = 0.45]{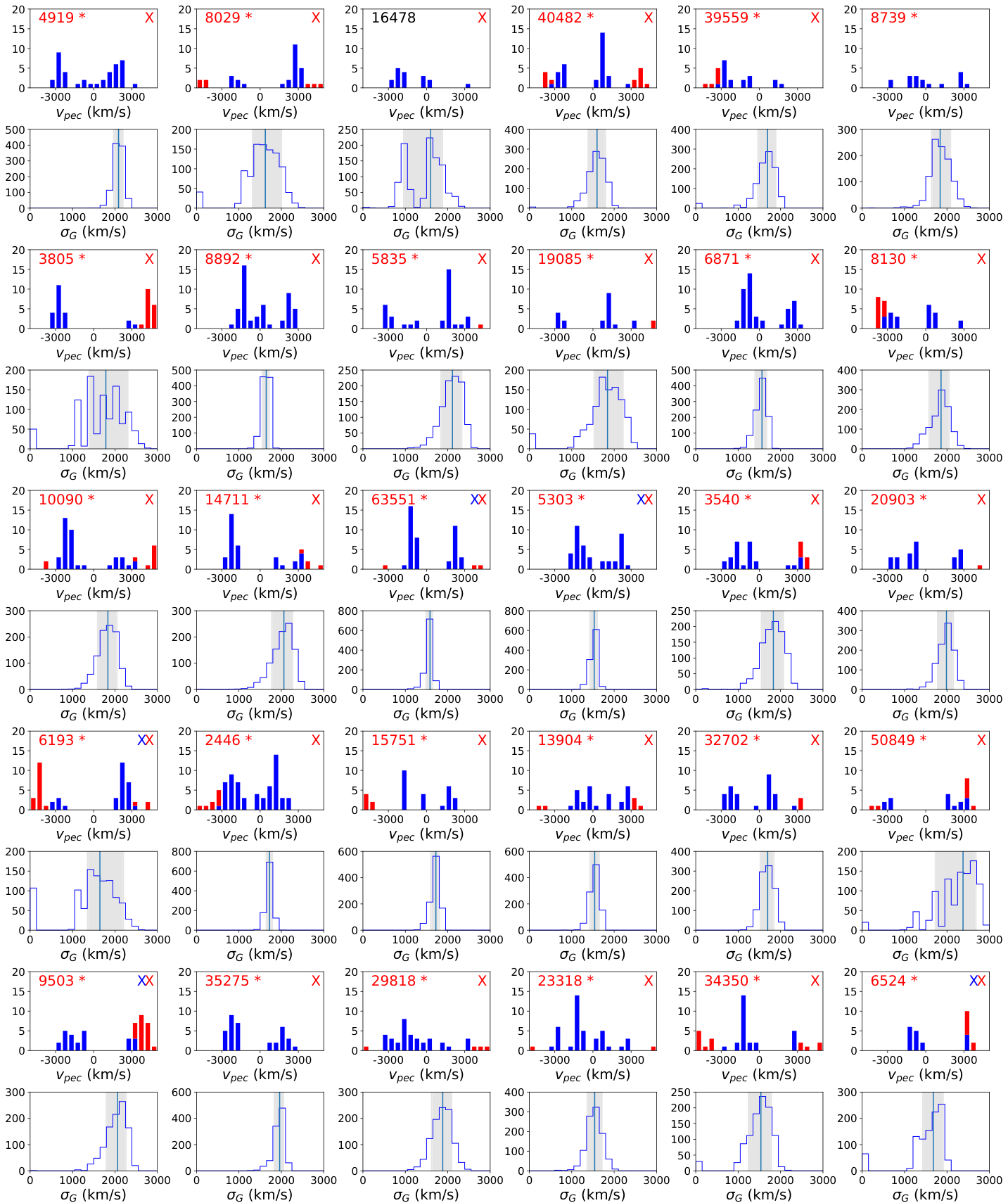}}
      \label{fig:veldistoutlier}
\end{figure}

\begin{figure}
    \ContinuedFloat
    \centering
    \subfloat{\includegraphics[scale = 0.45]{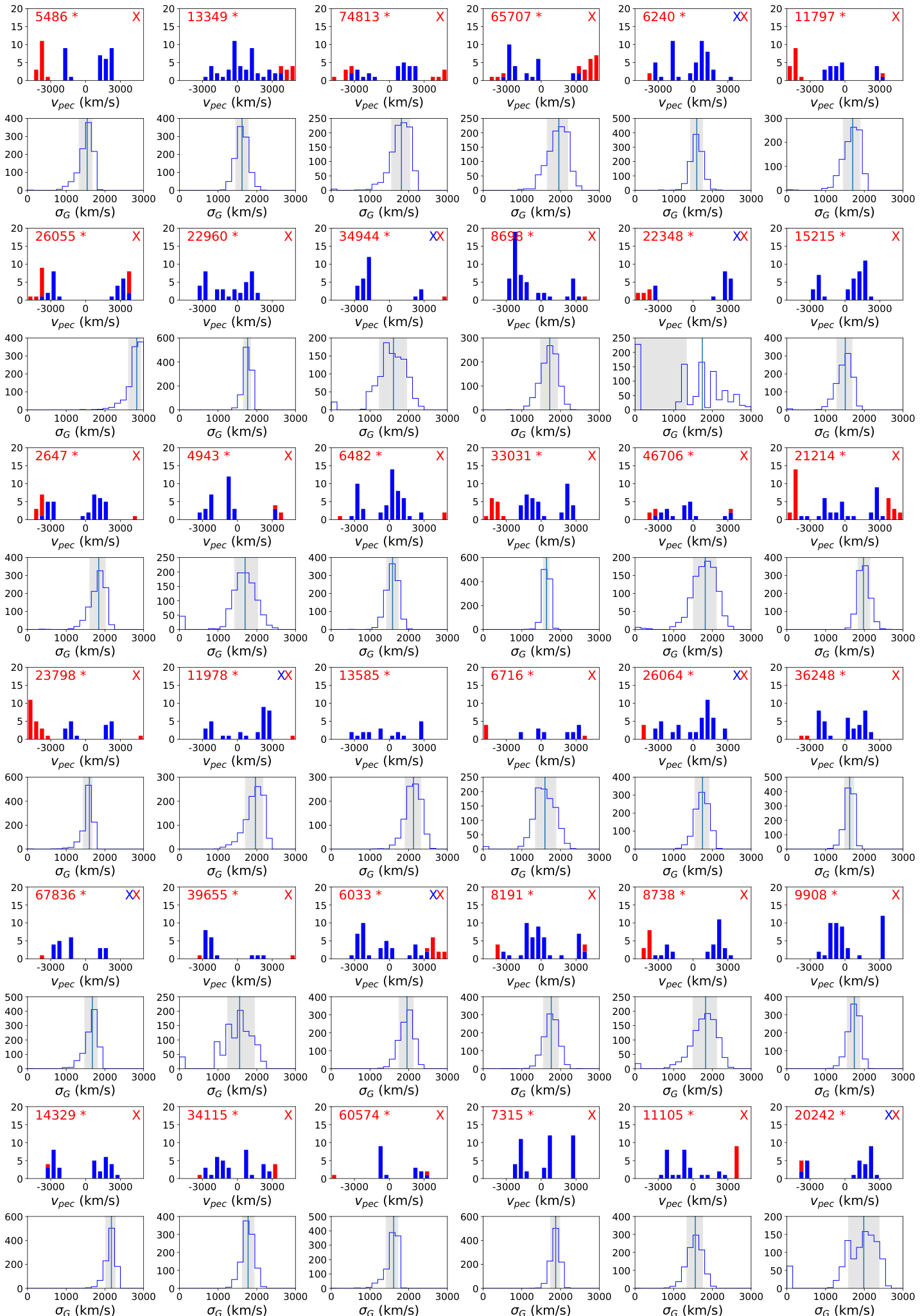}}
\end{figure}

\begin{figure}
    \ContinuedFloat
    \centering
    \subfloat{\includegraphics[scale = 0.45]{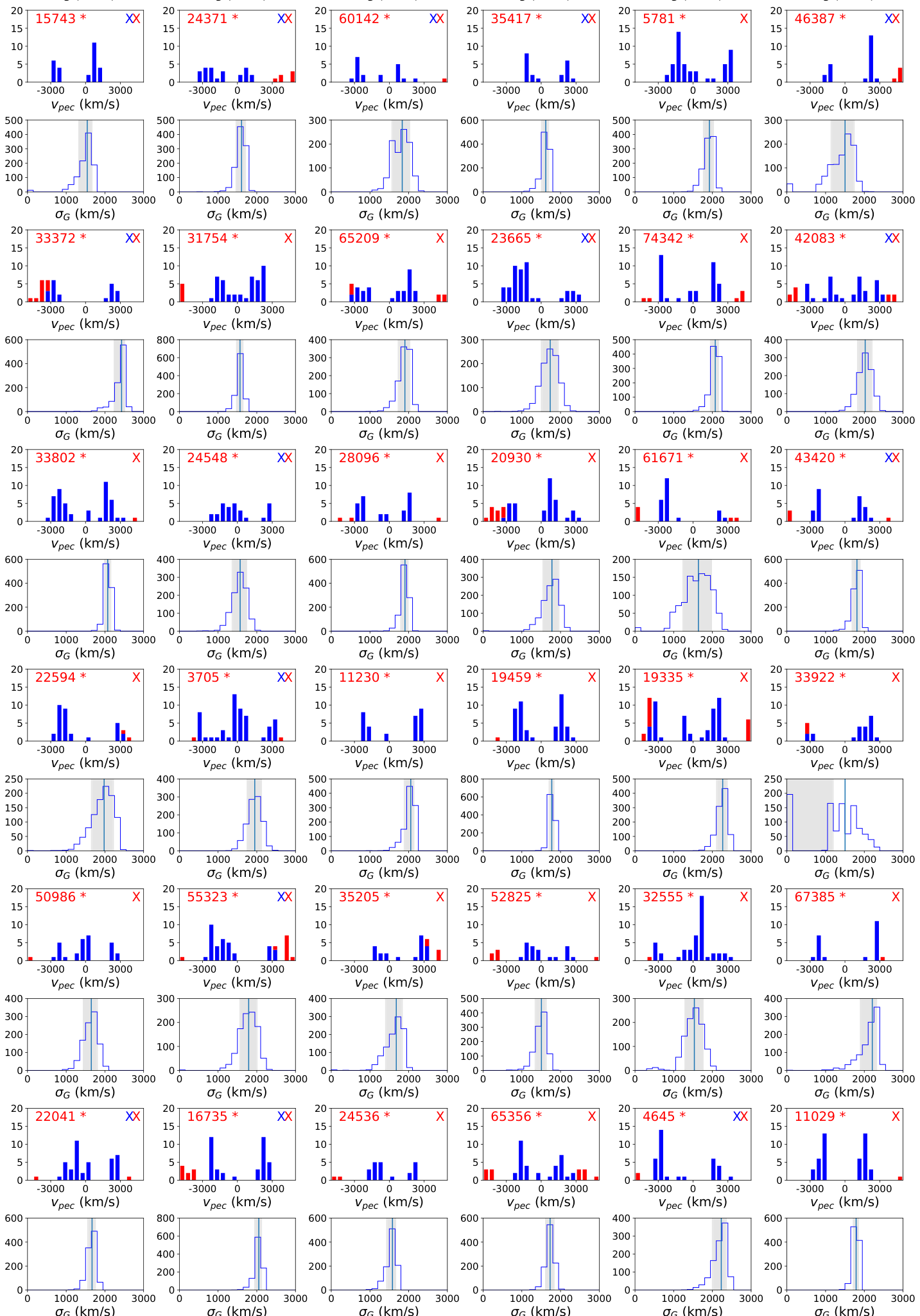}}
\end{figure}

\end{document}